\documentclass[useAMS,usenatbib]{mn2e}

\usepackage{graphicx}
\usepackage{caption}
\usepackage{amsmath}
\usepackage{mathrsfs}
\usepackage{natbib}
\usepackage{epstopdf}
\usepackage{float}

\title[A spectroscopic survey of Herbig Ae/Be stars with X-Shooter II]{A spectroscopic survey of Herbig Ae/Be stars with X-Shooter II: Accretion diagnostic lines. \thanks{Based on observations using the ESO Very Large Telescope, at Cerro Paranal
, under the observing program 084.C-0952A}}
\author[J.~R.~Fairlamb et al.]{J.~R.~Fairlamb$^{1}$\thanks{E-mail; john.fairlamb@gmail.com} R.~D.~Oudmaijer$^{1}$ I.~Mendigutia$^{1}$ J.~D.~Ilee$^{2}$ 
\newauthor
M.~E.~van den Ancker$^{3}$\\
$^{1}$School of Physics and Astronomy, EC Stoner Building, University of Leeds, Leeds, LS2 9JT, UK\\
$^{2}$ Institute of Astronomy, Madingley Road, Cambridge, CB3 0HA, UK\\
$^{3}$European Southern Observatory (ESO), Karl-Schwarzschild
-Str. 2, 85748 Garching, Germany 
}

\begin{document}

\def\km{{\rm\thinspace km}}
\def\pc{{\rm\thinspace pc}}
\def\kpc{{\rm\thinspace kpc}}
\def\s{{\rm\thinspace s}}
\def\au{{\rm\thinspace au}}
\def\yr{{\rm\thinspace yr}}
\def\Myr{{\rm\thinspace Myr}}
\def\myr{{\rm\thinspace Myr}}
\def\kmps{\hbox{${\rm\km\s^{-1}\,}$}}
\def\ewobs{\hbox{$\rm {\rm EW}_{\rm obs}$}}
\def\ewint{\hbox{$\rm {\rm EW}_{\rm int}$}}
\def\ewcor{\hbox{$\rm {\rm EW}_{\rm cor}$}}
\def\K{{\rm\thinspace K}}
\def\ergps{\hbox{${\rm\erg\s^{-1}\,}$}}
\def\Rsol{\hbox{${\rm\thinspace R_{\odot}}$}}
\def\Msol{\hbox{${\rm\thinspace M_{\odot}}$}}
\def\Lsol{\hbox{${\rm\thinspace L_{\odot}}$}}
\def\Macc{\hbox{${\thinspace \dot{M}_{\rm acc}}$}}
\def\Lacc{\hbox{${\thinspace L_{\rm acc}}$}}
\def\rsol{\hbox{${\rm\thinspace R_{\odot}}$}}
\def\msol{\hbox{${\rm\thinspace M_{\odot}}$}}
\def\lsol{\hbox{${\rm\thinspace L_{\odot}}$}}
\def\macc{\hbox{${\thinspace \dot{M}_{\rm acc}}$}}
\def\lacc{\hbox{${\thinspace L_{\rm acc}}$}}
\def\teff{\hbox{${\thinspace T_{\rm eff}}$}}
\def\Lline{\hbox{${\thinspace L_{\rm line}}$}}
\def\lline{\hbox{${\thinspace L_{\rm line}}$}}
\def\Fline{\hbox{${\thinspace F_{\rm line}}$}}
\def\fline{\hbox{${\thinspace F_{\rm line}}$}}
\def\micron{\hbox{${\thinspace {\rm \mu{}m}}$}}
\def\angstrom{\hbox{${\thinspace {\rm \AA}}$}}
\def\Angstrom{\hbox{${\thinspace {\rm \AA}}$}}
\def\kelvin{\hbox{${\thinspace {\rm K}}$}}
\def\Kelvin{\hbox{${\thinspace {\rm K}}$}}
\def\Msolpyr{\hbox{${\rm\Msol\yr^{-1}\,}$}}
\def\msolpyr{\hbox{${\rm\Msol\yr^{-1}\,}$}}

\newcommand{\foi}{[{O\,\sc{i}}]$_{\lambda6300}$}
\newcommand{\oisev}{{O\,\sc{i}}$_{\lambda{}7773}$}
\newcommand{\oieight}{{O\,\sc{i}}$_{\lambda{}8446}$}
\newcommand{\heifive}{{He\,\sc{i}}$_{\lambda{}5876}$}
\newcommand{\heioneo}{{He\,\sc{i}}$_{\lambda{}10830}$}
\newcommand{\oi}{{O\,\sc{i}}}
\newcommand{\caii}{{Ca\,\sc{ii}}}
\newcommand{\halpha}{H$\alpha$}
\newcommand{\hbeta}{H$\beta$}
\newcommand{\hgamma}{H$\gamma$}
\newcommand{\hdelta}{H$\delta$}
\newcommand{\brgamma}{Br$\gamma$}
\newcommand{\paalpha}{Pa$\alpha$}
\newcommand{\pabeta}{Pa$\beta$}
\newcommand{\pagamma}{Pa$\gamma$}
\newcommand{\padelta}{Pa$\delta$}
\newcommand{\paepsilon}{Pa$\epsilon$}

\def   \aj {{\rm { ApJ}}}
\def   \araa {{\rm { ARA\&{}A}}}
\def   \apj {{\rm {ApJ}}}
\def   \apjl {{\rm { ApJL}}}
\def   \apjs {{\rm { ApJS}}}
\def   \apss {{\rm { Ap\&{}SS}}}
\def   \aap {{\rm { A\&{}A}}}
\def   \aapr {{\rm { A&A Rev.}}}
\def   \aaps {{\rm { A\&{}AS}}}
\def   \astj {{\rm { ApJ}}}
\def   \baas {{\rm { BAAS}}}
\def   \memras {{\rm { MmRAS}}}
\def   \mnras {{\rm { MNRAS}}}
\def   \prl {{\rm { Phys. Rev. Lett.}}}
\def   \pasp {{\rm { PASP}}}
\def   \pasa {{\rm { PASA}}}
\def   \pasj {{\rm {PASJ}}}
\def   \nat {{\rm { Nature}}}
\def   \jqsrt{{\rm h{ J. Quant. Spectrosc. Rad. Trans.}}}
\def   \ssr {{\rm { Space Sci. Rev.}}}
\def   \arpc {{\rm h{Annu. Rev. Phys. Chem.}}}
\def   \rmxaa {{\rm h{Rev. Mex. Astron. Astr.}}}
\def   \pre {{\rm \emph{Phys. Rev. E}}}

\defcitealias{Calvet1998}{CG98}
\defcitealias{Fairlamb2015}{Paper I}

\date{Accepted 2016 October 12. Received 2016 October 10; in original form 2015 October 2}

\pagerange{\pageref{firstpage}--\pageref{lastpage}} \pubyear{2016}

\maketitle

\label{firstpage}

\begin{abstract}

The Herbig Ae/Be stars (HAeBes) allow an exploration of the properties of Pre-Main Sequence(PMS) stars above the low-mass range ($<2$\msol{}) and those bordering the high-mass range ($>8$\msol{}). This paper is the second in a series exploring accretion in 91 HAeBes with Very Large Telescope/X-shooter spectra. Equivalent width measurements are carried out on 32 different lines, spanning the UV to NIR, in order to obtain their line luminosities. The line luminosities were compared to accretion luminosities, which were determined directly from measurements of an UV-excess. When detected, emission lines always demonstrate a correlation with the accretion luminosity, regardless of detection frequency. The average relationship between accretion luminosity and line luminosity is found to be \lacc{}$ \propto$\lline{}$^{1.16 \pm 0.15}$. This is in agreement with the findings in Classical T Tauri stars, although the HAeBe relationship is generally steeper, particularly towards the Herbig Be mass range. Since all observed lines display a correlation with the accretion luminosity, all of them can be used as accretion tracers. This has increased the number of accretion diagnostic lines in HAeBes tenfold. However, questions still remain on the physical origin of each line, which may not be due to accretion.

\end{abstract}


\section{Introduction}
\label{sec:intro}

Herbig Ae/Be stars, HAeBes, bridge an important mass range between high and low mass stars; namely the intermediate masses of 2--8\msol{}. The HAeBes are pre-main sequence (PMS) stars that display properties such as an infra-red excess due to a circumstellar disc \citep{vandenAncker2000, Meeus2001}, and complex spectral line profiles \citep{Herbig1960, Finkenzeller1984, Hamann1992b}; akin to their solar-mass counterparts, the classical T Tauri stars, CTTs \citep[see][for a review]{Bertout1989}. The HAeBes serve as excellent observational targets which can further our understanding of star formation, as they are readily observable at optical wavelengths unlike the Massive Young Stellar Objects, MYSOs, which are still heavily embedded in their natal clouds \citep{Mottram2011}.

Theoretical models have long suggested that CTTs accrete via magnetospheric accretion, MA \citep{Ghosh1979, Uchida1985, Koenigl1991, Shu1994, Calvet1998}. In this scenario the circumstellar disc is truncated at a given radius from the star by the stellar magnetic field. The magnetic field lines can then funnel disc material onto the star, at approximately free-fall velocities, via an accretion column. This in-falling material shocks the photosphere and gives rise to hotspots on the stellar surface. The result is an excess of ultra-violet (UV) energy which is detectable in addition to the regular photospheric emission. This phenomenon has been observed in numerous CTTs to date \citep{Calvet1998, Gullbring1998, Gullbring2000, Ingleby2013}, and has been shown to be present in HAeBes \citep{Garrison1978, Muzerolle2004, Donehew2011, Mendigutia2011b, Mendigutia2013, Fairlamb2015}. Provided the stellar parameters are known, the accretion luminosity can be straight-forwardly converted into a mass accretion rate, \macc{}, one of the key astrophysical parameters in star formation. 

It has also been shown that the accretion luminosity (\lacc{}) in CTTs, derived from an UV-excess or line veiling, is correlated with the luminosity of emission lines (\lline{}). This appears to hold true for a large number of lines, such as Br$\gamma$, H$\alpha$, \foi{}, and the {Ca\,\sc{ii}} IR-triplet \citep{Muzerolle1998b, Calvet2004, Dahm2008, Herczeg2008, Rigliaco2012, Alcala2014}. To date, the luminosity relationships for CTTs have largely been shown to hold for HAeBes too, with only minor changes in intercept and gradient \citep{Mendigutia2011b,Mendigutia2013}. 
This means that emission lines can be used as powerful accretion diagnostics and provide a method of inferring \macc{} for stars where direct methods of measurement are difficult or impossible e.g. the UV-excess is often used to measure \macc{} directly, but UV-observations are more difficult to obtain and interpret than simple emission lines; emission lines are also available at a variety of different wavelengths making them a preferred choice over UV observations in highly extinct targets. Therefore, by establishing correlations between the accretion luminosity and line luminosity for a large series of lines, possibilities to infer accretion rates when using data sets with a limited wavelength range are opened. 

Changes in the accretion mechanism are suspected to be taking place towards the early-Be stars, as evidenced by spectropolarimetric observations \citep{Vink2002, Vink2005}. It was also thought that the \lacc{} and \lline{} relationship was breaking down towards HBes by \citet{Donehew2011}. However, the observed change in the relationship can be explained by the adopted MA model which was for a single cool HAe star. In fact, the observed trends between \lacc{} and \lline{} have been explained by \citet{Mendigutia2015} to be a consequence of the luminosity of the star. Therefore, many observed relationships may not actually be directly due to accretion. However, any observed relationships can still be used to infer an accretion rate, meaning that they remain a valuable tool in this field. Therefore, a large scale investigation into these relationships for HAeBes is required.

In our previous paper in this series \citep[][hereafter \citetalias{Fairlamb2015}]{Fairlamb2015}, we used spectroscopic data from X-Shooter, covering the UV, optical and near-infrared wavelengths, to determine the stellar parameters of a large sample of 91 HAeBe stars.  Out of this sample, it was possible to determine the UV-excess, and thereby the accretion luminosity and mass accretion rates, for 62 of them. The large wavelength range of the data allows for the detection and measurement of many emission lines; observed simultaneously with the UV-excess. The large sample is therefore ideal to revisit the hitherto published line luminosities and accretion luminosity relationships, along with an investigation into new accretion diagnostics.

The overall aim of this paper is to provide the measurement of numerous spectral lines across the entire sample of 91 HAeBes. The strengths of the lines are investigated, along with a determinations of line luminosities based on previous stellar properties and photometry. These are compared against the accretion rates determined in \citetalias{Fairlamb2015} in order to provide a critical assessment of accretion luminosity versus line luminosity relationships in HAeBes.

This paper is organised as follows. Section \ref{sec:data} provides
details on the sample, data reduction, and techniques of emission line
measurements and line luminosity calculations. Section
\ref{sec:results} presents the measurements of the lines, calculated line luminosities, and the resulting correlations with the accretion luminosity. Analysis of these relationships and the presentation of new accretion diagnostics are provided in this section too. Finally, Section \ref{sec:disc_and_conc} provides a summary of the results and the concluding remarks.

\section{Sample, Observations, Data Reduction, and Measurements}
\label{sec:data}

The sample, observations and basic data reduction processes are detailed in paper I. We give a brief summary of these here, while we discuss the correction for telluric absorption lines and go into detail on the measurement of line strengths. The stars were selected from the catalogues of HAeBe stars published by \citet{The1994} and \citet{Vieira2003}. The main criterion to be included the sample was their observability in the southern hemisphere during the (southern) summer semester. In total the sample contains 91 stars, of which 37 have a mass greater than 3\msol{}, which we choose to classify as the Herbig Be stars. The remaining stars are all HAes, with perhaps a few intermediate mass CTTs which border the mass boundary. 

The observations were carried out with X-Shooter mounted on the ESO's VLT, UT3, Paranal, Chile in the 2009-2010 season \citep{Vernet2011}. The instrument provides complete coverage from 0.3-2.4 $\mu$m over three separate arms; the arms are split into the following: the UVB arm, 3000--5600~\AA; the VIS arm, 5500--10200~${\rm \AA}$; and the NIR arm, 10200-24800~${\rm \AA}$. The observations were taken with the smallest possible slit widths resulting in a spectral resolution of 10,000 - 18,000. 

The integration times were such that signal-to-noise ratios of 100-200 were typical. The bias subtraction, flat fielding and wavelength calibration were all performed using the ESO pipeline reduction (Modigliani et al. 2010). When available, data were taken from the recent phase III release with pipeline version 2.3.12.  For the small number where this was not the case, the reduced data from paper I (pipeline version 0.9.7) are taken. In general, the quality of the reduced data in both cases is similar, except in the NIR arm where the SNR is better in the phase III data.

The spectra were not flux calibrated as a matter of course, and instead we use published photometry of all our sources when computing the final line luminosities (see \citetalias{Fairlamb2015} for details of photometry sources). Since the typical errors in the flux determination from spectrophometric data amount to around 20\%, this choice is warranted for objects that vary less than 20\% (or roughly 0.2 magnitude). It has been shown in PMS-stars that the variability amplitude decreases with increasing wavelength e.g. for a change in the V-band of 0.4~mag changes in the K-band are generally less than 0.2~mag \citep{Eiroa2001,Eiroa2002}. Spot-checks on our targets indicate that the variability is indeed at most of this order on average, validating the choice of using existing photometry. Due to the large sample used any objects with large variability changes, that could produce anomalous results, are likely to be identified or their effects diluted by the sheer number of other targets used.

\subsection{Correction for telluric absorption lines}
\label{sec:data:tell_correction}

Before measurements of the lines are undertaken, telluric absorption
features in the spectra must be removed. This is particularly
necessary for the majority of the observable emission lines in the
near-infrared (NIR) arm, but also for a few cases in the VIS arm
covering the optical.

A telluric standard star can be used for such a correction, but these stars contain their own spectral features, which also need to be corrected for. Since many of the lines observed are hydrogen recombination lines, which are commonly seen in absorption in telluric standard stars, correction is instead performed using the ESO Molecfit\footnote{http://www.eso.org/sci/software/pipelines/skytools/molecfit} telluric removal tool \citep{Smette2015, Kausch2015}. There are a few additional issues that contributed to rejecting the telluric standard and using Molecfit instead: airmass could differ by large amounts (mostly uncommon), and short exposure times of the tellurics (on average the SNR was half that of the targets).
Molecfit provides an atmospheric transmission spectrum using a radiative transfer code based upon a combination of the observatory conditions at the time and the features observed in the target spectra.
Removal of the lines using both Molecfit and telluric standards were compared and it was found that using Molecfit yielded better quality spectra over using the telluric standards; the SNR of the latter contributes to a poorer reduction.

Therefore, Molecfit correction is performed on all spectral orders in each arm that contains telluric line features, and will be used for the remainder of this paper \citep[see also][where Molecfit is used in the same way to study the CO overtone emission at 2.3$\mu$m]{Ilee2014}.

\subsection{Measuring emission line strengths}
\label{sec:data:ew_measuring}

After telluric corrections, line strengths are recorded by measuring
the equivalent widths, EW, of the lines after normalising the spectra
of each star to the continuum level. These direct measurements of the EW
will be referred to as \ewobs{}. In order to obtain the true line
strengths of the emission lines \ewobs{} needs to be corrected for both the underlying photospheric
absorption, \ewint{}, and the presence of any excess continuum
emission, which would dilute the underlying absorption.

\begin{figure}
\centering
\includegraphics[trim=0.5cm 0.75cm 0.5cm 0.75cm, width=\linewidth]{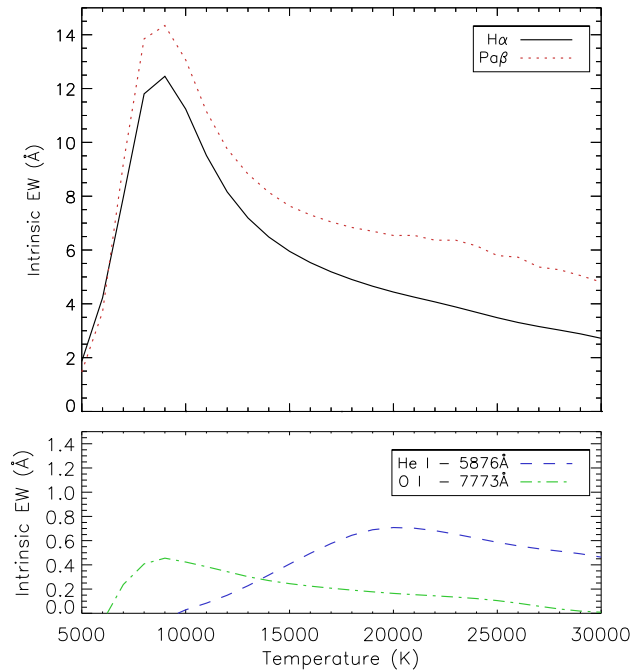}
\caption{ The two panels above shown how \ewint{} varies as a function of temperature. More importantly it demonstrates that the variation is significantly different between different elements e.g. Hydrogen absorption peaks at $\sim 9000$~K, while Helium absorption peaks at $\sim 20000$~K.} 
\label{fig:ew_curves}
\end{figure}

The photospheric absorption depends upon the temperature, surface
gravity, and metallicity of the stars. The temperature and surface gravity of each star were determined in \citetalias{Fairlamb2015} and are used to select an
appropriate spectral model atmosphere, from which the absorption line
strengths are measured. In this work the Munari set of Kurucz-Castelli
models, KC-models, are used for lines with $\lambda <
10050$\angstrom{}, due to their 1\angstrom{} sampling
\citep{Munari2005}; a metalicity of [M/H]=0 is adopted. For longer wavelengths additional models are required, the older KC-models are used here computed by \citet{Kurucz1993,
Castelli2004}. For consistency, the same region of measurement used
for \ewobs{} is used for measuring \ewint{}. 
Figure \ref{fig:ew_curves} provides an example of the intrinsic
absorption strength as a function of temperature
for some of the lines (log(g) is fixed in this example). 

The net emission, which is stronger than the straightforward measurement of \ewobs{} would imply, can now be
corrected for the intrinsic absorption.  The corrected equivalent
width, \ewcor{}, is calculated as follows: 
\begin{equation} 
{\rm EW_{cor}} = {\rm EW_{obs}} - 10^{-0.4\Delta m_\lambda}{\rm EW_{int}} 
\label{eqn:ew_cor} 
\end{equation} 
where $\Delta
m_\lambda$ is the strength of the continuum excess at the wavelength in question, measured in magnitudes. It is the difference between the observed, dereddened magnitude and the intrinsic magnitude of a
star of the same spectral type.

\begin{figure}
\centering
\includegraphics[trim=0.55cm 0.7cm 0.55cm 0.6cm, width=\linewidth]{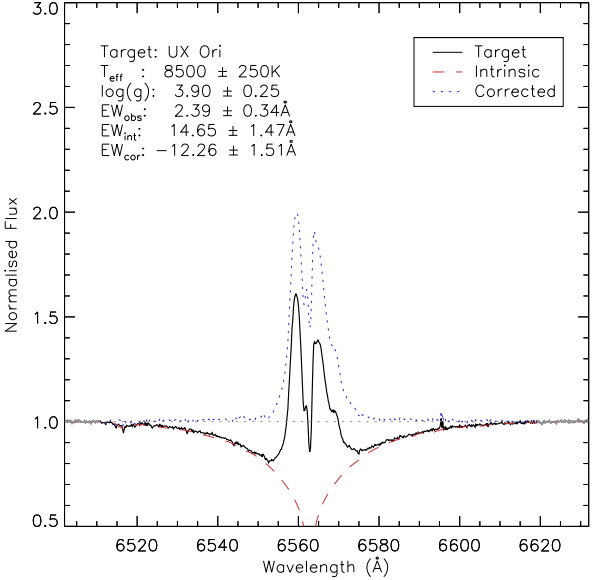}
\caption{ The \halpha{} line for UX Ori is displayed here; the original spectra is shown in black. Also plotted here are: the predicted intrinsic spectra, red-dashed; the corrected spectra, blue-dotted; the continuum level, grey-dashed; and finally the left and right continuum regions which are used for normalising the lines, solid-grey. The left-hand legend details the properties of UX Ori. The figure demonstrates that despite clear emission a positive \ewobs{} can be measured. Therefore it is also important to assess \ewint{} when examining the strength of lines.} 
\label{fig:ux_ori}
\end{figure}

Figure \ref{fig:ux_ori} demonstrates the importance of correcting for intrinsic absorption. In the figure it can be seen that UX Ori definitely has emission present. However, when measuring \ewobs{} a positive value is obtained i.e. an absorption line. Correcting for \ewint{} makes \ewcor{} negative, and therefore in emission. The benefit of this approach is that the result encompasses all of the true emission.

At near-infrared wavelengths excess emission is also present, due
to thermal emission from circumstellar dust heated by the central
star. 
i.e. there is a non-zero $\Delta m_\lambda$ component, for the majority of stars, because
of the IR-excess. In order to obtain $\Delta m_\lambda{}$ the intrinsic magnitude is first obtained by
scaling a KC-model of the same stellar parameters to match the
dereddened visible photometry of the star and measuring the flux from the model at the
wavelength in question (the compiled photometry is
listed in paper I). To avoid underestimating the continuum flux,
which can be affected by the underlying absorption in the model, an interpolation
of the continuum regions either side of the line is used to derive the
continuum flux. This is then converted into a magnitude to give the
expected intrinsic magnitude (since this is at an arbitrary wavelength no convolutions with pass-bands are involved). Next, this must be compared against the observed dereddened
magnitude. This makes use of the $JHK$ photometric data from 2MASS
\citep{Cutri2003, Skrutskie2006}. The photometry is dereddened and
converted into a flux using the $A_V$ values determined in
\citetalias{Fairlamb2015}, and then an interpolation is made between the photometric points either side of the wavelength in question (and converted back into a magnitude).
With both the observed and expected intrinsic magnitudes obtained,
$\Delta m_\lambda$ is found via the difference; \ewcor{}
can now be calculated.

\subsection{Error budget}
\label{sec:data:error_budget}

A consideration of errors needs to be made in all the above steps. Let us first start with the line measurement itself. Since \ewobs{} is a
numerical calculation over a wavelength region, the error will be the
extent of this region, $\Delta\lambda$ divided by the SNR of the continuum: $ \sigma_{\rm EW
obs}= \Delta\lambda /{\rm SNR}$. The error on \ewint{}
is also complicated by stellar parameters, metallicity, and the
adopted set of models. For this reason a generous error of 10\% is
adopted for all \ewint{} measured from the Munari set of
KC-Models; an error of 20\% is adopted for the NIR lines measured from the lower resolution models. Errors in $\Delta
m_\lambda$ are low, due to the quality of the $JHK$ photometry,
and
the relatively low errors in the stellar parameters. 
Since the EW errors dominate over $\Delta
m_\lambda$ only the \ewobs{} and \ewint{} errors are considered in \ewcor{}.
With \ewcor{} established, a line flux, \fline{}, can now be
calculated by \fline{} = \ewcor{}$ \times F_\lambda$, where
$F_\lambda$ is the continuum flux corresponding to the central
wavelength of the line. For all lines with line centres $<1$\micron{},
$F_\lambda$ is calculated from a KC-model in the same way the
intrinsic magnitude was calculated earlier (but omitting the final
magnitude conversion). For the lines with line centres $>1$\micron{},
an interpolation of the dereddened $JHK$ photometry is performed in
the same manner as was done for obtaining the IR-excess magnitude
earlier, except a further step is required of converting this into a
flux.

Expanding upon this, a line luminosity, \lline{} ,is also
calculated by: \lline{}$ = 4 \pi
D^2 $\fline{}, where $D$ is the distance to star (the values used are complied in
\citetalias{Fairlamb2015}). The \lline{} value takes into account both
errors on the distance and the stellar parameters
used to calculate \fline{}. A complete list of errors on \lline{} are included in the online version of table \ref{tab:ew_sample}.

\section{Results}
\label{sec:results}

In total, \ewobs{}, \ewint{}, \ewcor{} and the corresponding line luminosities
are calculated for a set of 32 different lines. The lines were selected based on being
observed previously in the literature, or by having a high
detection rate throughout the sample; preferably both. These lines span the full wavelength range covered. To put the number of lines into perspective, we note that lines previously identified in the literature as accretion diagnostics for HAeBes stars are limited to H$\alpha$, [O{\sc i}]6300 and Br$\gamma$ \citep{Donehew2011,Mendigutia2011b}; hence the number of lines investigated in this manner has increased tenfold. 

The EW, \fline{}, and \lline{} data for the H$\alpha$ line are provided in Table \ref{tab:ew_sample}. The table contains this information for all of the observed targets. A further 31 lines have been measured and are available online.

The following sections investigate the empirical relationship between the line
luminosities and the accretion luminosities, along with a comparison against CTTs.

\begin{table*}
\centering
\begin{minipage}{160mm}
\caption[EW table]{This table provides the EW measurements for the H$\alpha$ line in all objects. A full version of this table containing all 32 lines is available online. It is worth noting that the $\Delta m_\lambda$ column contains no values for this particular line as no correction for an excess is required; an IR-excess correction is required for lines $>$1\micron{}. In column 6, containing the line flux, the `E$-x$' factor denotes $\times10^x$.}  
\label{tab:ew_sample}
\begin{tabular}{l rrr rcc}
\hline
\multicolumn{7}{c}{Details of the H$\alpha$ line} \\
\hline
Name  & \multicolumn{1}{c}{\ewobs{}} & \multicolumn{1}{c}{\ewint{}} & \multicolumn{1}{c}{$\Delta m_\lambda$} & \multicolumn{1}{c}{\ewcor{}} & \fline{} &  log(\lline{}) \\
      & \multicolumn{1}{c}{(\angstrom{})} & \multicolumn{1}{c}{(\angstrom{})} & \multicolumn{1}{c}{(mag)} &  \multicolumn{1}{c}{(\angstrom{})} & (Wm$^{-2}$\Angstrom{}$^{-1}$) & [\lsol] \\
\hline

UX Ori          &     2.39 $\pm$    0.34 &    14.65 $\pm$    1.47 &     - &   -12.26 $\pm$    1.51 &     3.27 ($\pm$  0.40) E -15 &     -1.43 $\pm$   0.10 \\ 
PDS 174         &   -54.19 $\pm$    1.34 &     6.46 $\pm$    0.65 &     - &   -60.65 $\pm$    1.49 &     2.33 ($\pm$  0.06) E -14 &     -0.03 $\pm$   0.09 \\ 
V1012 Ori       &     4.74 $\pm$    0.55 &    16.36 $\pm$    1.64 &     - &   -11.62 $\pm$    1.73 &     1.42 ($\pm$  0.21) E -15 &     -2.06 $\pm$   0.11 \\ 
HD 34282        &     4.18 $\pm$    0.40 &    16.23 $\pm$    1.62 &     - &   -12.05 $\pm$    1.67 &     3.03 ($\pm$  0.42) E -15 &     -1.90 $\pm$   0.11 \\ 
HD 287823       &    10.18 $\pm$    0.50 &    15.58 $\pm$    1.56 &     - &    -5.40 $\pm$    1.64 &     1.64 ($\pm$  0.50) E -15 &     -2.23 $\pm$   0.16 \\ 
HD 287841       &     8.36 $\pm$    0.43 &    13.22 $\pm$    1.32 &     - &    -4.86 $\pm$    1.39 &     9.64 ($\pm$  2.75) E -16 &     -2.46 $\pm$   0.15 \\ 
HD 290409       &     0.35 $\pm$    0.55 &    15.00 $\pm$    1.50 &     - &   -14.65 $\pm$    1.60 &     3.18 ($\pm$  0.35) E -15 &     -1.58 $\pm$   0.10 \\ 
HD 35929        &     1.54 $\pm$    0.69 &     8.69 $\pm$    0.87 &     - &    -7.15 $\pm$    1.11 &     1.09 ($\pm$  0.17) E -14 &     -1.35 $\pm$   0.11 \\ 
HD 290500       &    -1.46 $\pm$    0.25 &    13.14 $\pm$    1.31 &     - &   -14.60 $\pm$    1.33 &     1.26 ($\pm$  0.11) E -15 &     -1.04 $\pm$   0.10 \\ 
HD 244314       &   -14.04 $\pm$    0.61 &    15.54 $\pm$    1.55 &     - &   -29.58 $\pm$    1.67 &     6.91 ($\pm$  0.39) E -15 &     -1.38 $\pm$   0.09 \\ 
HK Ori          &   -61.86 $\pm$    0.72 &    15.79 $\pm$    1.58 &     - &   -77.65 $\pm$    1.74 &     1.51 ($\pm$  0.03) E -14 &     -1.04 $\pm$   0.09 \\ 
HD 244604       &     2.32 $\pm$    0.51 &    14.79 $\pm$    1.48 &     - &   -12.47 $\pm$    1.57 &     5.72 ($\pm$  0.72) E -15 &     -1.46 $\pm$   0.10 \\ 
UY Ori          &     4.98 $\pm$    0.26 &    15.28 $\pm$    1.53 &     - &   -10.30 $\pm$    1.55 &     4.81 ($\pm$  0.72) E -16 &     -1.80 $\pm$   0.11 \\ 
HD 245185       &   -13.56 $\pm$    0.50 &    14.48 $\pm$    1.45 &     - &   -28.04 $\pm$    1.53 &     6.54 ($\pm$  0.36) E -15 &     -1.26 $\pm$   0.09 \\ 
T Ori           &    -4.15 $\pm$    0.43 &    13.07 $\pm$    1.31 &     - &   -17.22 $\pm$    1.38 &     1.04 ($\pm$  0.08) E -14 &     -0.74 $\pm$   0.09 \\ 
V380 Ori        &   -81.88 $\pm$    0.48 &    13.61 $\pm$    1.36 &     - &   -95.49 $\pm$    1.44 &     9.74 ($\pm$  0.15) E -14 &     -0.48 $\pm$   0.09 \\ 
HD 37258        &    -0.33 $\pm$    0.39 &    15.00 $\pm$    1.50 &     - &   -15.33 $\pm$    1.55 &     4.88 ($\pm$  0.49) E -15 &     -1.56 $\pm$   0.10 \\ 
HD 290770       &   -24.08 $\pm$    0.35 &    12.98 $\pm$    1.30 &     - &   -37.06 $\pm$    1.35 &     1.57 ($\pm$  0.06) E -14 &     -1.02 $\pm$   0.09 \\ 
BF Ori          &    -0.02 $\pm$    0.46 &    14.70 $\pm$    1.47 &     - &   -14.72 $\pm$    1.54 &     5.35 ($\pm$  0.56) E -15 &     -1.36 $\pm$   0.10 \\ 
HD 37357        &     4.76 $\pm$    0.41 &    14.64 $\pm$    1.46 &     - &    -9.88 $\pm$    1.52 &     6.32 ($\pm$  0.97) E -15 &     -1.63 $\pm$   0.11 \\ 
HD 290764       &    -2.38 $\pm$    0.43 &    13.31 $\pm$    1.33 &     - &   -15.69 $\pm$    1.40 &     5.03 ($\pm$  0.45) E -15 &     -1.46 $\pm$   0.10 \\ 
HD 37411        &    -1.07 $\pm$    0.47 &    15.55 $\pm$    1.56 &     - &   -16.62 $\pm$    1.63 &     5.29 ($\pm$  0.52) E -15 &     -1.67 $\pm$   0.10 \\ 
V599 Ori        &     1.69 $\pm$    0.57 &    13.35 $\pm$    1.34 &     - &   -11.66 $\pm$    1.46 &     6.48 ($\pm$  0.81) E -15 &     -1.28 $\pm$   0.10 \\ 
V350 Ori        &     3.09 $\pm$    0.33 &    15.66 $\pm$    1.57 &     - &   -12.57 $\pm$    1.60 &     2.52 ($\pm$  0.32) E -15 &     -1.69 $\pm$   0.10 \\ 
HD 250550       &   -48.83 $\pm$    0.39 &    10.03 $\pm$    1.00 &     - &   -58.86 $\pm$    1.07 &     2.01 ($\pm$  0.04) E -14 &     -0.22 $\pm$   0.09 \\ 
V791 Mon        &   -88.08 $\pm$    0.25 &     7.94 $\pm$    0.79 &     - &   -96.02 $\pm$    0.83 &     4.22 ($\pm$  0.04) E -14 &     -0.26 $\pm$   0.09 \\ 
PDS 124         &   -13.83 $\pm$    1.37 &    14.13 $\pm$    1.41 &     - &   -27.96 $\pm$    1.97 &     2.02 ($\pm$  0.14) E -15 &     -1.30 $\pm$   0.09 \\ 
LkHa 339        &    -5.39 $\pm$    1.37 &    12.98 $\pm$    1.30 &     - &   -18.37 $\pm$    1.89 &     4.24 ($\pm$  0.44) E -15 &     -1.33 $\pm$   0.10 \\ 
VY Mon          &   -17.81 $\pm$    1.76 &     8.45 $\pm$    0.84 &     - &   -26.26 $\pm$    1.95 &     6.94 ($\pm$  0.52) E -14 &     -0.38 $\pm$   0.09 \\ 
R Mon           &  -114.51 $\pm$    1.76 &     9.35 $\pm$    0.93 &     - &  -123.86 $\pm$    1.99 &     4.22 ($\pm$  0.07) E -14 &     -0.07 $\pm$   0.09 \\ 
V590 Mon        &   -60.10 $\pm$    0.52 &     9.63 $\pm$    0.96 &     - &   -69.73 $\pm$    1.09 &     3.46 ($\pm$  0.05) E -15 &     -0.49 $\pm$   0.09 \\ 
PDS 24          &   -25.47 $\pm$    1.79 &    12.98 $\pm$    1.30 &     - &   -38.45 $\pm$    2.21 &     1.17 ($\pm$  0.07) E -15 &     -1.00 $\pm$   0.09 \\ 
PDS 130         &   -31.17 $\pm$    0.70 &    11.41 $\pm$    1.14 &     - &   -42.58 $\pm$    1.34 &     2.75 ($\pm$  0.09) E -15 &     -0.58 $\pm$   0.09 \\ 
PDS 229N        &     7.41 $\pm$    0.82 &     9.63 $\pm$    0.96 &     - &    -2.22 $\pm$    1.26 &     1.72 ($\pm$  0.97) E -16 &     -1.99 $\pm$   0.26 \\ 
GU CMa          &   -14.86 $\pm$    0.48 &     4.75 $\pm$    0.47 &     - &   -19.61 $\pm$    0.67 &     1.61 ($\pm$  0.06) E -13 &      0.15 $\pm$   0.09 \\ 
HT CMa          &   -20.94 $\pm$    0.35 &    11.87 $\pm$    1.19 &     - &   -32.81 $\pm$    1.24 &     1.75 ($\pm$  0.07) E -15 &     -0.84 $\pm$   0.09 \\ 
Z CMa           &   -63.55 $\pm$    0.99 &     9.97 $\pm$    1.00 &     - &   -73.52 $\pm$    1.41 &     7.72 ($\pm$  0.15) E -13 &      1.43 $\pm$   0.09 \\ 
HU CMa          &   -52.00 $\pm$    0.40 &     9.09 $\pm$    0.91 &     - &   -61.09 $\pm$    0.99 &     6.64 ($\pm$  0.11) E -15 &     -0.50 $\pm$   0.09 \\ 
HD 53367        &    -7.62 $\pm$    0.52 &     4.02 $\pm$    0.40 &     - &   -11.64 $\pm$    0.66 &     2.15 ($\pm$  0.12) E -13 &     -0.11 $\pm$   0.09 \\ 
PDS 241         &    -8.36 $\pm$    0.29 &     4.21 $\pm$    0.42 &     - &   -12.57 $\pm$    0.51 &     4.32 ($\pm$  0.17) E -15 &      0.06 $\pm$   0.09 \\ 
NX Pup          &   -37.01 $\pm$    0.45 &     8.74 $\pm$    0.87 &     - &   -45.75 $\pm$    0.98 &     1.75 ($\pm$  0.04) E -14 &     -1.04 $\pm$   0.09 \\ 
PDS 27          &   -73.20 $\pm$    0.73 &     4.40 $\pm$    0.44 &     - &   -77.60 $\pm$    0.85 &     1.08 ($\pm$  0.01) E -13 &      1.53 $\pm$   0.09 \\ 
PDS 133         &  -103.11 $\pm$    3.91 &     7.84 $\pm$    0.78 &     - &  -110.95 $\pm$    3.99 &     4.99 ($\pm$  0.18) E -15 &     -0.01 $\pm$   0.09 \\ 
HD 59319        &     5.86 $\pm$    6.14 &     7.13 $\pm$    0.71 &     - &    -1.27 $\pm$    6.18 & $<$ 1.19             E -15 & $<$ -1.26             \\ 
PDS 134         &   -12.22 $\pm$    0.41 &     5.98 $\pm$    0.60 &     - &   -18.20 $\pm$    0.73 &     1.58 ($\pm$  0.06) E -15 &      0.20 $\pm$   0.09 \\ 
HD 68695        &     0.48 $\pm$    0.48 &    16.51 $\pm$    1.65 &     - &   -16.03 $\pm$    1.72 &     4.18 ($\pm$  0.45) E -15 &     -1.81 $\pm$   0.10 \\ 
HD 72106        &     8.78 $\pm$    0.63 &    14.60 $\pm$    1.46 &     - &    -5.82 $\pm$    1.59 &     4.77 ($\pm$  1.30) E -15 &     -1.69 $\pm$   0.15 \\ 
TYC 8581-2002-1 &    11.05 $\pm$    0.42 &    13.61 $\pm$    1.36 &     - &    -2.56 $\pm$    1.42 &     3.50 ($\pm$  1.95) E -16 &     -2.05 $\pm$   0.26 \\ 
PDS 33          &    -3.10 $\pm$    0.48 &    15.83 $\pm$    1.58 &     - &   -18.93 $\pm$    1.65 &     7.90 ($\pm$  0.69) E -16 &     -1.67 $\pm$   0.09 \\ 
HD 76534        &   -11.00 $\pm$    0.34 &     5.84 $\pm$    0.58 &     - &   -16.84 $\pm$    0.67 &     3.63 ($\pm$  0.15) E -14 &     -0.44 $\pm$   0.09 \\ 
PDS 281         &     4.30 $\pm$    0.49 &     5.42 $\pm$    0.54 &     - &    -1.12 $\pm$    0.73 &     3.81 ($\pm$  2.48) E -15 &     -0.98 $\pm$   0.30 \\ 
PDS 286         &   -26.84 $\pm$    0.42 &     3.93 $\pm$    0.39 &     - &   -30.77 $\pm$    0.57 &     2.75 ($\pm$  0.05) E -13 &      0.37 $\pm$   0.09 \\ 
PDS 297         &     7.42 $\pm$    0.41 &    11.36 $\pm$    1.14 &     - &    -3.94 $\pm$    1.21 &     2.80 ($\pm$  0.86) E -16 &     -1.73 $\pm$   0.16 \\ 
HD 85567        &   -50.25 $\pm$    0.47 &     6.76 $\pm$    0.68 &     - &   -57.01 $\pm$    0.83 &     1.09 ($\pm$  0.02) E -13 &      0.45 $\pm$   0.09 \\ 
HD 87403        &     6.79 $\pm$    0.63 &     9.76 $\pm$    0.98 &     - &    -2.97 $\pm$    1.17 &     1.27 ($\pm$  0.50) E -15 &     -0.89 $\pm$   0.19 \\ 
PDS 37          &  -119.89 $\pm$    0.47 &     3.87 $\pm$    0.39 &     - &  -123.76 $\pm$    0.61 &     2.53 ($\pm$  0.01) E -13 &      2.17 $\pm$   0.09 \\ 
HS 305298       &     0.01 $\pm$    0.41 &     3.25 $\pm$    0.32 &     - &    -3.24 $\pm$    0.52 &     9.93 ($\pm$  1.59) E -16 &     -0.45 $\pm$   0.11 \\ 
HD 94509        &   -16.84 $\pm$    0.53 &     6.33 $\pm$    0.63 &     - &   -23.17 $\pm$    0.82 &     1.09 ($\pm$  0.04) E -14 &      0.82 $\pm$   0.09 \\ 

\hline
\end{tabular}
\end{minipage}
\end{table*}

\begin{table*}
\centering
\begin{minipage}{160mm}
\contcaption{}
\begin{tabular}{lrrrrcc}
\hline
\multicolumn{7}{c}{Details of the XXXXXX line} \\
\hline
Name  & \multicolumn{1}{c}{\ewobs{}} & \multicolumn{1}{c}{\ewint{}} & \multicolumn{1}{c}{$\Delta m_\lambda$} & \multicolumn{1}{c}{\ewcor{}} & \fline{} &  log(\lline{}) \\
      & \multicolumn{1}{c}{(\angstrom{})} & \multicolumn{1}{c}{(\angstrom{})} & \multicolumn{1}{c}{(mag)} &  \multicolumn{1}{c}{(\angstrom{})} & (Wm$^{-2}$\angstrom{}$^{-1}$) & [\lsol] \\
\hline

HD 95881        &   -12.53 $\pm$    0.39 &     9.35 $\pm$    0.94 &     - &   -21.88 $\pm$    1.02 &     2.60 ($\pm$  0.12) E -14 &      0.13 $\pm$   0.09 \\ 
HD 96042        &     3.20 $\pm$    0.80 &     3.95 $\pm$    0.39 &     - &    -0.75 $\pm$    0.89 & $<$ 1.27             E -15 & $<$ -0.90             \\ 
HD 97048        &   -24.43 $\pm$    0.34 &    13.54 $\pm$    1.35 &     - &   -37.97 $\pm$    1.39 &     8.06 ($\pm$  0.30) E -14 &     -1.13 $\pm$   0.09 \\ 
HD 98922        &   -14.64 $\pm$    0.44 &    10.04 $\pm$    1.00 &     - &   -24.68 $\pm$    1.09 &     1.16 ($\pm$  0.05) E -13 &     -0.36 $\pm$   0.09 \\ 
HD 100453       &     5.27 $\pm$    0.77 &    10.23 $\pm$    1.02 &     - &    -4.96 $\pm$    1.28 &     9.50 ($\pm$  2.45) E -15 &     -2.35 $\pm$   0.14 \\ 
HD 100546       &   -23.96 $\pm$    0.42 &    15.50 $\pm$    1.55 &     - &   -39.46 $\pm$    1.61 &     1.76 ($\pm$  0.07) E -13 &     -1.29 $\pm$   0.09 \\ 
HD 101412       &    -0.15 $\pm$    0.69 &    15.28 $\pm$    1.53 &     - &   -15.43 $\pm$    1.68 &     8.38 ($\pm$  0.91) E -15 &     -1.62 $\pm$   0.10 \\ 
PDS 344         &   -22.23 $\pm$    0.45 &     7.80 $\pm$    0.78 &     - &   -30.03 $\pm$    0.90 &     7.69 ($\pm$  0.23) E -16 &     -0.74 $\pm$   0.09 \\ 
HD 104237       &   -13.74 $\pm$    0.98 &    13.75 $\pm$    1.38 &     - &   -27.49 $\pm$    1.69 &     1.61 ($\pm$  0.10) E -13 &     -1.18 $\pm$   0.09 \\ 
V1028 Cen       &  -101.61 $\pm$    0.52 &     7.01 $\pm$    0.70 &     - &  -108.62 $\pm$    0.87 &     2.22 ($\pm$  0.02) E -14 &      0.37 $\pm$   0.09 \\ 
PDS 361S        &    -3.97 $\pm$    0.45 &     5.35 $\pm$    0.53 &     - &    -9.32 $\pm$    0.70 &     8.10 ($\pm$  0.60) E -16 &     -0.31 $\pm$   0.09 \\ 
HD 114981       &    -7.60 $\pm$    0.46 &     5.64 $\pm$    0.56 &     - &   -13.24 $\pm$    0.72 &     3.37 ($\pm$  0.18) E -14 &     -0.06 $\pm$   0.09 \\ 
PDS 364         &   -78.38 $\pm$    0.74 &     9.63 $\pm$    0.96 &     - &   -88.01 $\pm$    1.21 &     4.40 ($\pm$  0.06) E -15 &     -0.39 $\pm$   0.09 \\ 
PDS 69          &   -69.42 $\pm$    1.30 &     7.00 $\pm$    0.70 &     - &   -76.42 $\pm$    1.48 &     8.40 ($\pm$  0.16) E -14 &      0.02 $\pm$   0.09 \\ 
DG Cir          &   -47.98 $\pm$    0.83 &    13.10 $\pm$    1.31 &     - &   -61.08 $\pm$    1.55 &     6.27 ($\pm$  0.16) E -15 &     -1.00 $\pm$   0.09 \\ 
HD 132947       &    10.38 $\pm$    0.33 &    11.96 $\pm$    1.20 &     - &    -1.58 $\pm$    1.24 &     9.04 ($\pm$  7.12) E -16 &     -2.04 $\pm$   0.35 \\ 
HD 135344B      &    -4.60 $\pm$    0.75 &     5.66 $\pm$    0.57 &     - &   -10.26 $\pm$    0.94 &     1.35 ($\pm$  0.12) E -14 &     -2.08 $\pm$   0.10 \\ 
HD 139614       &     0.48 $\pm$    0.61 &    13.26 $\pm$    1.33 &     - &   -12.78 $\pm$    1.46 &     1.33 ($\pm$  0.15) E -14 &     -2.09 $\pm$   0.10 \\ 
PDS 144S        &   -16.12 $\pm$    0.68 &    13.07 $\pm$    1.31 &     - &   -29.19 $\pm$    1.48 &     9.49 ($\pm$  0.48) E -16 &     -1.53 $\pm$   0.09 \\ 
HD 141569       &     5.13 $\pm$    0.57 &    15.55 $\pm$    1.56 &     - &   -10.42 $\pm$    1.66 &     3.36 ($\pm$  0.54) E -14 &     -1.88 $\pm$   0.11 \\ 
HD 141926       &   -43.48 $\pm$    0.43 &     3.40 $\pm$    0.34 &     - &   -46.88 $\pm$    0.55 &     3.00 ($\pm$  0.04) E -13 &      1.17 $\pm$   0.09 \\ 
HD 142666       &     5.14 $\pm$    0.68 &    11.70 $\pm$    1.17 &     - &    -6.56 $\pm$    1.35 &     9.17 ($\pm$  1.89) E -15 &     -2.22 $\pm$   0.12 \\ 
HD 142527       &    -6.95 $\pm$    0.98 &     6.18 $\pm$    0.62 &     - &   -13.13 $\pm$    1.16 &     1.91 ($\pm$  0.17) E -14 &     -1.93 $\pm$   0.10 \\ 
HD 144432       &    -0.98 $\pm$    0.68 &    11.65 $\pm$    1.17 &     - &   -12.63 $\pm$    1.35 &     1.76 ($\pm$  0.19) E -14 &     -1.85 $\pm$   0.10 \\ 
HD 144668       &    -8.41 $\pm$    0.45 &    14.12 $\pm$    1.41 &     - &   -22.53 $\pm$    1.48 &     1.40 ($\pm$  0.09) E -13 &     -0.95 $\pm$   0.09 \\ 
HD 145718       &     8.34 $\pm$    0.52 &    14.53 $\pm$    1.45 &     - &    -6.19 $\pm$    1.54 &     6.86 ($\pm$  1.71) E -15 &     -2.34 $\pm$   0.14 \\ 
PDS 415N        &     3.10 $\pm$    1.30 &     5.04 $\pm$    0.50 &     - &    -1.94 $\pm$    1.39 &     2.45 ($\pm$  1.76) E -16 &     -3.53 $\pm$   0.32 \\ 
HD 150193       &    -5.59 $\pm$    0.57 &    16.07 $\pm$    1.61 &     - &   -21.66 $\pm$    1.71 &     6.21 ($\pm$  0.49) E -14 &     -1.55 $\pm$   0.09 \\ 
AK Sco          &    -0.82 $\pm$    0.95 &     5.06 $\pm$    0.51 &     - &    -5.88 $\pm$    1.08 &     4.81 ($\pm$  0.88) E -15 &     -2.60 $\pm$   0.12 \\ 
PDS 431         &     1.27 $\pm$    0.50 &    10.49 $\pm$    1.05 &     - &    -9.22 $\pm$    1.16 &     4.41 ($\pm$  0.56) E -16 &     -0.94 $\pm$   0.10 \\ 
KK Oph          &   -25.15 $\pm$    0.31 &    16.36 $\pm$    1.64 &     - &   -41.51 $\pm$    1.67 &     1.28 ($\pm$  0.05) E -14 &     -1.50 $\pm$   0.09 \\ 
HD 163296       &    -3.63 $\pm$    0.36 &    16.02 $\pm$    1.60 &     - &   -19.65 $\pm$    1.64 &     7.82 ($\pm$  0.65) E -14 &     -1.60 $\pm$   0.09 \\ 
MWC 297         &  -590.00 $\pm$    0.90 &     4.52 $\pm$    0.45 &     - &  -594.52 $\pm$    1.01 &     4.68 ($\pm$  0.01) E -11 &      1.63 $\pm$   0.09 \\ 

\hline
\end{tabular}
\end{minipage}
\end{table*}

\subsection{Accretion luminosity - line luminosity relationships}
\label{sec:results:relationships}
 
\begin{figure*}
\centering
\includegraphics[trim=0.75cm 0.25cm 0.5cm 0.5cm, width=0.95\linewidth]{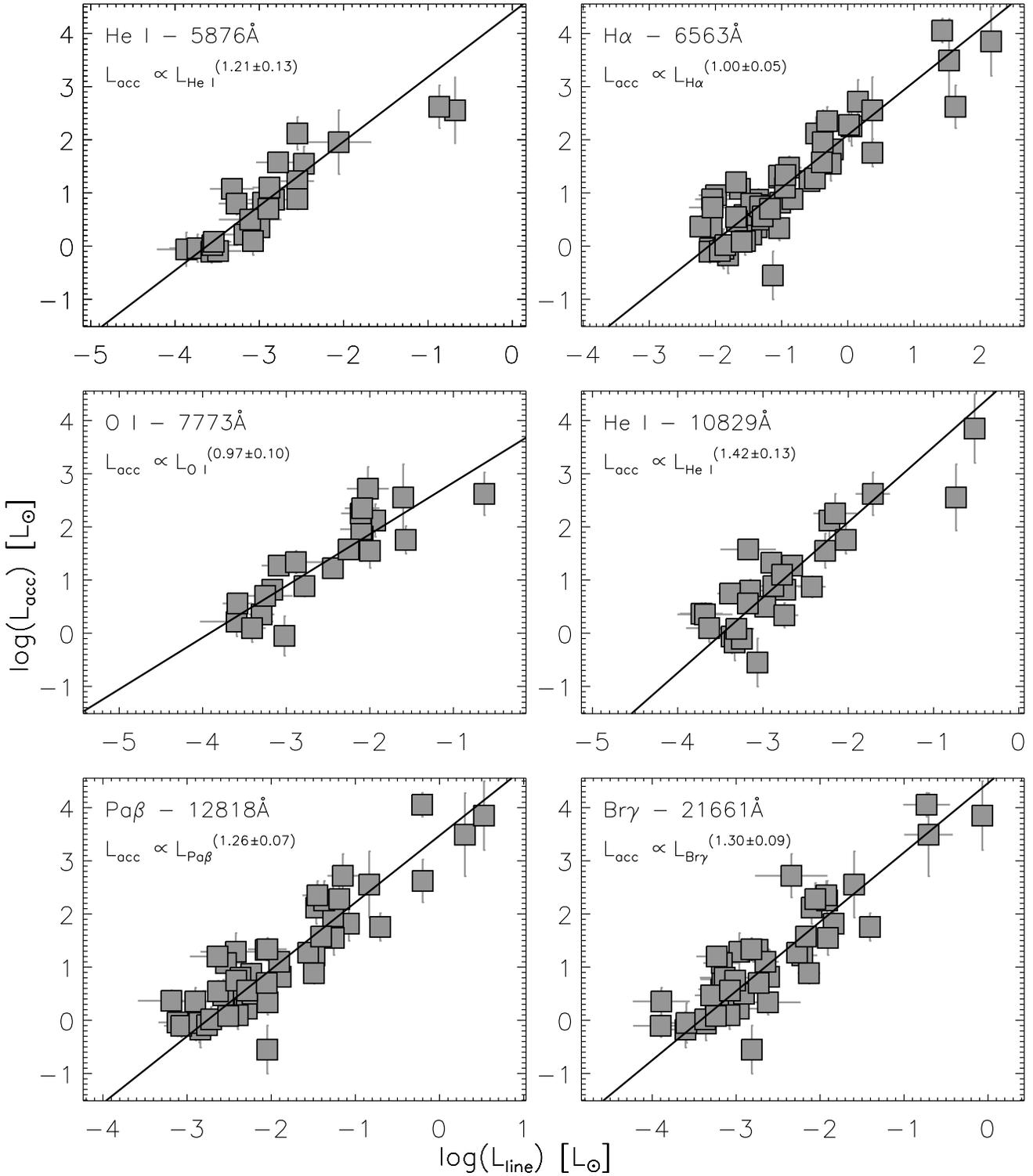}
\caption{ Relationships are displayed between the accretion luminosities determined in \citetalias{Fairlamb2015} and the line luminosities determined in this work. Each panel shows only the HAeBes where both a clear UV-excess and clear emission line detection are made, shown as grey squares;a best-fit to them is shown as a solid-black line (the equation for the line is given below the line descriptor in the top-left of each panel). The relationships are all comparable to each other with the average exponent of each relationship being $\sim 1.19 \pm 0.18$.}
\label{fig:lacc_vs_lline}
\end{figure*}

In Figure \ref{fig:lacc_vs_lline} the relationship between the accretion luminosity \lacc{}, derived for 62 stars using the UV-excess, is shown plotted against the luminosity of six different emission lines. The selected lines are \heifive{}, \halpha{}, \oisev{}, \heioneo{}, \pabeta{}, and \brgamma{} as they cover a large wavelength range and a few different elemental species. Correlations for other lines are available in the on-line version of this paper.

The line luminosity--accretion luminosity relationship can be approximated by a power law, and has been in numerous works on CTTs and HAeBes \citep{Muzerolle1998b, Calvet2004, Dahm2008, Herczeg2008, Mendigutia2011b, Rigliaco2012}, where :

\begin{equation}
{\rm log} \left( \frac{L_{\rm acc}}{\lsol{}} \right) = A + B \times {\rm log} \left( \frac{L_{\rm line}}{\lsol{}} \right)
\label{eqn:lacc_vs_lline}
\end{equation}
where $A$ and $B$ are constants describing the intercept and slope respectively. Such a relationship is obtained across all lines for the Herbig Ae/Be stars; the slopes and intercepts are presented in Table \ref{tab:lacc_lum}. The best-fits were calculated using the IDL \textit{mpfitexy} package, which factors in errors in both the $x$ and $y$ axis.
Figure \ref{fig:lacc_vs_lline_append}, in the Appendix, displays the correlations for all lines observed; Table \ref{tab:lacc_lum} provides details of the correlations. Where possible the findings of \citet{Alcala2014} on a set of CTTs are shown in these figures for comparison. The same relationship between the line and accretion luminosities observed for CTTs, and reported previously in HAeBes, extends to other lines as well. Indeed, this trend is essentially observed for \textit{all} lines. The strong correlation between line luminosity and accretion luminosity is the reason why emission lines are a useful tool for easily determining accretion luminosities (and in turn determining the mass accretion rate).

A comparison with previous determinations of the slope and intercept derived in the literature for Herbig Ae/Be stars is made here. Specifically the \halpha{}, \foi{}, and \brgamma{}  relationships from \citet{Mendigutia2011b} are assessed. It is found that the relationships obtained here agree with previous work; they lie within the errorbars of each other. The main difference here is that the errors, particularly on the intercept, are reduced here. This is presumably due to the increased sample size and simultaneity of the observations of the UV excess and lines.

The relationship for \brgamma{} in this paper agrees well with \citet{Mendigutia2011b}. Both of these relationships do not agree with the breakdown in \brgamma{} relationship implied by \citet{Donehew2011}. This can be attributed to the MA models used in the latter being associated with a relatively low mass HAeBe star, producing lower accretion luminosities than expected; their model was based upon UX Ori from \citet{Muzerolle2004}, while this work focuses on multiple models from \citet{
Mendigutia2011b} and \citetalias{Fairlamb2015}. Therefore, we can say that there remains a \lacc{}-\lline{} relationship for HAeBes.

\subsection{Comparison with T Tauri stars}
\label{sec:results:ctts_compare}

The data on CTTs from \citet{Alcala2014} for the lines in common with
this study are also included for comparison, along with their best fit
in Figure \ref{fig:lacc_vs_lline_append}. Overall, the plots appear to show a qualitative
agreement in trends between the HAeBes and the CTTs, where \lacc{} is
increasing proportionally to \lline{}, generally a 1:1 correlation.  The placement of the CTTs and HAeBes in the plots often
agree with each where they transition between each other, suggesting similarity between the two groups. An exception is the Ca II triplet where there is an offset, likely due to blending with the Paschen series. The best-fit relationship to the HAeBes is generally steeper than it is for the extrapolated CTTs best-fit, suggesting a possible difference in the origin of the line luminosity. However, this is not true in all cases and the average of all relationships are in agreement with each other. The slope of the \lacc{} vs. \lline{} relationship, the $B$ value, is
found to have an average of $\bar{B} = 1.16 \pm 0.15 $ in the
HAeBes, while for the CTTs it is
$\bar{B} = 1.08 \pm 0.08$. 
The values of the best-fit relationships for the CTTs and the HAeBes are provided in Table \ref{tab:lacc_lum}.

\begin{table*}
 \centering
 
\begin{minipage}{145mm}
  \caption{This table details the number of emission lines detected for each line (Column 4) and the number of these used in determining \lacc{}-\lline{} relationships (Column 5). The nubmer of lines used is less as they must have a corresponding UV-excess to be included. The best-fitting slopes and intercepts to the luminosity relationships plotted in Figures \ref{fig:lacc_vs_lline} and \ref{fig:lacc_vs_lline_append} are provided in columns 6 and 7. The best-fit parameters for the CTTs presented by \citet{Alcala2014} are provided in the final two columns for comparison.}
  \label{tab:lacc_lum}
  \begin{tabular}{ c cl cc cc cc}
  
  \hline
  Line & $\lambda$ & Line & No. & No. & \multicolumn{2}{c}{This work} & \multicolumn{2}{c}{\citet{Alcala2014}} \\
  No. & ($\rm \AA$) &  & Detections & Used & $A \pm \sigma_A$ & $B$ $\pm \sigma_{\rm B}$ & A $\pm \sigma_A$ & $B \pm \sigma_B$ \\
  \hline

 1 &  3797 & H(10-2) & 42 & 30 &  3.04 $\pm$  0.14 &  1.14 $\pm$  0.08 &  2.58 $\pm$  0.27 &  1.00 $\pm$  0.05 \\
 2 &  3835 & H(9-2) & 24 & 16 &  2.91 $\pm$  0.14 &  1.06 $\pm$  0.10 &  2.53 $\pm$  0.27 &  1.01 $\pm$  0.05 \\
 3 &  3889 & H(8-2) & 22 & 18 &  2.96 $\pm$  0.16 &  1.17 $\pm$  0.11 &  2.55 $\pm$  0.29 &  1.04 $\pm$  0.06 \\
 4 &  4102 & Hdelta & 23 & 14 &  2.65 $\pm$  0.13 &  1.14 $\pm$  0.10 &  2.50 $\pm$  0.28 &  1.06 $\pm$  0.06 \\
 5 &  4340 & Hgamma & 37 & 25 &  2.51 $\pm$  0.10 &  1.10 $\pm$  0.09 &  2.50 $\pm$  0.25 &  1.09 $\pm$  0.05 \\
 6 &  4861 & Hbeta & 81 & 51 &  2.60 $\pm$  0.09 &  1.24 $\pm$  0.07 &  2.31 $\pm$  0.23 &  1.11 $\pm$  0.05 \\
 7 &  5876 & He I & 31 & 24 &  4.39 $\pm$  0.38 &  1.21 $\pm$  0.13 &  3.51 $\pm$  0.30 &  1.13 $\pm$  0.06 \\
 8 &  6300 & [O I] & 48 & 28 &  3.84 $\pm$  0.16 &  0.94 $\pm$  0.06 &   -   $\pm$   -   &   -   $\pm$   -   \\
 9 &  6563 & Halpha & 89 & 54 &  2.09 $\pm$  0.06 &  1.00 $\pm$  0.05 &  1.50 $\pm$  0.26 &  1.12 $\pm$  0.07 \\
10 &  7773 & O I & 29 & 21 &  3.80 $\pm$  0.26 &  0.97 $\pm$  0.10 &  3.91 $\pm$  0.51 &  1.16 $\pm$  0.09 \\
11 &  8446 & O I & 57 & 37 &  3.61 $\pm$  0.14 &  0.90 $\pm$  0.05 &  3.06 $\pm$  0.90 &  1.06 $\pm$  0.18 \\
12 &  8498 & Ca II & 41 & 29 &  3.50 $\pm$  0.14 &  0.91 $\pm$  0.06 &  2.18 $\pm$  0.38 &  0.95 $\pm$  0.07 \\
13 &  8542 & Ca II & 34 & 26 &  3.62 $\pm$  0.15 &  1.04 $\pm$  0.07 &  2.13 $\pm$  0.42 &  0.95 $\pm$  0.08 \\
14 &  8598 & Pa(14-3) & 74 & 49 &  3.88 $\pm$  0.14 &  1.13 $\pm$  0.06 &   -   $\pm$   -   &   -   $\pm$   -   \\
15 &  8662 & Ca II & 79 & 52 &  3.45 $\pm$  0.12 &  1.08 $\pm$  0.05 &  2.20 $\pm$  0.43 &  0.95 $\pm$  0.09 \\
16 &  8750 & Pa(12-3) & 89 & 55 &  3.87 $\pm$  0.14 &  1.34 $\pm$  0.06 &   -   $\pm$   -   &   -   $\pm$   -   \\
17 &  8863 & Pa(11-3) & 78 & 51 &  3.81 $\pm$  0.15 &  1.29 $\pm$  0.07 &   -   $\pm$   -   &   -   $\pm$   -   \\
18 &  9015 & Pa(10-3) & 87 & 54 &  3.81 $\pm$  0.15 &  1.43 $\pm$  0.07 &  2.99 $\pm$  0.49 &  1.03 $\pm$  0.09 \\
19 &  9229 & Pa(9-3) & 73 & 46 &  3.72 $\pm$  0.15 &  1.31 $\pm$  0.07 &  3.40 $\pm$  0.47 &  1.13 $\pm$  0.09 \\
20 &  9546 & Paepsilon & 86 & 54 &  3.75 $\pm$  0.14 &  1.38 $\pm$  0.07 &  3.19 $\pm$  0.58 &  1.11 $\pm$  0.12 \\
21 & 10049 & Padelta & 65 & 43 &  4.01 $\pm$  0.17 &  1.26 $\pm$  0.07 &  3.33 $\pm$  0.47 &  1.18 $\pm$  0.10 \\
22 & 10829 & He I & 40 & 27 &  4.92 $\pm$  0.38 &  1.42 $\pm$  0.13 &  2.62 $\pm$  0.57 &  1.11 $\pm$  0.12 \\
23 & 10938 & Pagamma & 71 & 46 &  3.76 $\pm$  0.16 &  1.26 $\pm$  0.07 &  3.17 $\pm$  0.31 &  1.18 $\pm$  0.06 \\
24 & 12818 & Pabeta & 78 & 50 &  3.47 $\pm$  0.13 &  1.26 $\pm$  0.07 &  2.45 $\pm$  0.39 &  1.04 $\pm$  0.08 \\
25 & 15439 & Br(17-4) & 27 & 17 &  4.25 $\pm$  0.27 &  0.94 $\pm$  0.11 &   -   $\pm$   -   &   -   $\pm$   -   \\
26 & 15556 & Br(16-4) & 41 & 26 &  4.35 $\pm$  0.21 &  1.05 $\pm$  0.08 &   -   $\pm$   -   &   -   $\pm$   -   \\
27 & 15701 & Br(15-4) & 37 & 25 &  4.33 $\pm$  0.21 &  1.06 $\pm$  0.08 &   -   $\pm$   -   &   -   $\pm$   -   \\
28 & 15880 & Br(14-4) & 33 & 23 &  4.27 $\pm$  0.23 &  1.09 $\pm$  0.09 &   -   $\pm$   -   &   -   $\pm$   -   \\
29 & 16109 & Br(13-4) & 55 & 40 &  4.41 $\pm$  0.20 &  1.21 $\pm$  0.08 &   -   $\pm$   -   &   -   $\pm$   -   \\
30 & 16407 & Br(12-4) & 44 & 30 &  4.37 $\pm$  0.22 &  1.22 $\pm$  0.09 &   -   $\pm$   -   &   -   $\pm$   -   \\
31 & 16806 & Br(11-4) & 61 & 44 &  4.60 $\pm$  0.21 &  1.32 $\pm$  0.08 &   -   $\pm$   -   &   -   $\pm$   -   \\
32 & 21661 & Brgamma & 69 & 43 &  4.46 $\pm$  0.23 &  1.30 $\pm$  0.09 &  3.60 $\pm$  0.38 &  1.16 $\pm$  0.07 \\
  \hline
  \end{tabular}
\end{minipage}

\end{table*}

Overall, it appears that the relationship between \lacc{} and
\lline{}, for all lines, is well correlated in HAeBes. This confirms the predictions of \citet{Mendigutia2015} that the relationship is a mathematical one tied to the luminosity stars rather than the actual strengths of the lines
The slope of the relationships obtained here are slightly enhanced over the set of CTTs analysed by
\citet{Alcala2014}. A small level of caution should be noted if using the relationships of the \caii{}
triplet due to Paschen blending which may be causing the offset observed between HAeBes and CTTs. Another note of caution is advised when using the He I and O I lines, as their complex line profiles suggest various origins -- which may or may not be associated with
accretion. The complexities of their profiles are likely the cause of the low detection rate of these lines, as multiple absorption and emission
components can contribute to the line equivalent width. An exploration of the line profiles will be presented in a future paper by the authors.

\subsection{Emission lines as accretion diagnostics:}
\label{sec:results:diagnostics}

Since all of the emission lines appear to be correlated with the accretion luminosity it is worth discussing which lines serve as the most sensitive tracers of accretion i.e. most readily detected (and therefore possibly associated directly with accretion). That way the best lines can be prioritised in future observations.

\begin{table*}
\centering

\begin{minipage}{145mm}

\caption[Line detection statistics]{ This table details all of the measured emission lines, along with their respective number of detections. The final four columns denote the four categories into which each star of the sample can belong to for a given line; full descriptions are provided in the text. `Emis' denotes emission line, and $\Delta D_B$ denotes the UV-excess. }
\label{tab:yes_no}

\begin{tabular}{ccll cccc}
\hline
Line   &  $\lambda$  & Line & Emission & \multicolumn{4}{c}{No. of stars which match the criteria} \\
Number & ($\rm \AA$) &         & Lines  & Emis - Y  &  Emis - N & Emis - Y & Emis - N \\
       &             &         & Detected & $\Delta D_B$ - Y  &  $\Delta D_B$ - Y & $\Delta D_B$ - N & $\Delta D_B$ - N \\

\hline

1 &   3797 &  H(10-2) & 42 & 36 & 27 &  6 & 22 \\ 
2 &   3835 &  H(9-2) & 24 & 22 & 41 &  2 & 26 \\ 
3 &   3889 &  H(8-2) & 22 & 22 & 41 &  0 & 28 \\ 
4 &   4102 &  H$\delta$ & 23 & 20 & 43 &  3 & 25 \\ 
5 &   4340 &  H$\gamma$ & 37 & 31 & 32 &  6 & 22 \\ 
6 &   4861 &  H$\beta$ & 81 & 58 &  5 & 23 &  5 \\ 
7 &   5876 &  He I & 31 & 28 & 35 &  3 & 25 \\ 
8 &   6300 &  [O I] & 48 & 34 & 29 & 14 & 14 \\ 
9 &   6563 &  H$\alpha$ & 89 & 61 &  2 & 28 &  0 \\ 
10 &   7773 &  O I & 29 & 25 & 38 &  4 & 24 \\ 
11 &   8446 &  O I & 57 & 43 & 20 & 14 & 14 \\ 
12 &   8498 &  Ca II & 41 & 35 & 28 &  6 & 22 \\ 
13 &   8542 &  Ca II & 34 & 32 & 31 &  2 & 26 \\ 
14 &   8598 &  Pa(14-3) & 74 & 56 &  7 & 18 & 10 \\ 
15 &   8662 &  Ca II & 79 & 58 &  5 & 21 &  7 \\ 
16 &   8750 &  Pa(12-3) & 89 & 62 &  1 & 27 &  1 \\ 
17 &   8863 &  Pa(11-3) & 78 & 57 &  6 & 21 &  7 \\ 
18 &   9015 &  Pa(10-3) & 87 & 61 &  2 & 26 &  2 \\ 
19 &   9229 &  Pa(9-3) & 73 & 52 & 11 & 21 &  7 \\ 
20 &   9546 &  Pa$\epsilon$ & 86 & 61 &  2 & 25 &  3 \\ 
21 &  10049 &  Pa$\delta$ & 65 & 48 & 15 & 17 & 11 \\ 
22 &  10829 &  He I & 40 & 32 & 31 &  8 & 20 \\ 
23 &  10938 &  Pa$\gamma$ & 71 & 52 & 11 & 19 &  9 \\ 
24 &  12818 &  Pa$\beta$ & 78 & 57 &  6 & 21 &  7 \\ 
25 &  15439 &  Br(17-4) & 27 & 23 & 40 &  4 & 24 \\ 
26 &  15556 &  Br(16-4) & 41 & 32 & 31 &  9 & 19 \\ 
27 &  15701 &  Br(15-4) & 37 & 31 & 32 &  6 & 22 \\ 
28 &  15880 &  Br(14-4) & 33 & 28 & 35 &  5 & 23 \\ 
29 &  16109 &  Br(13-4) & 55 & 46 & 17 &  9 & 19 \\ 
30 &  16407 &  Br(12-4) & 44 & 36 & 27 &  8 & 20 \\ 
31 &  16806 &  Br(11-4) & 61 & 50 & 13 & 11 & 17 \\ 
32 &  21661 &  Br$\gamma$ & 69 & 50 & 13 & 19 &  9 \\

\hline
\end{tabular}  
  
\end{minipage}
\end{table*}

In Table \ref{tab:yes_no} all 32 lines are presented along with
details of the number of emission detections. The other columns
of the table are split into different categories based upon emission
detection and UV-excess detection. The total accretion rate detection
is defined as all of the HAeBes for which an UV-excess was clearly
detected in \citetalias{Fairlamb2015} -- 62 stars match
this criteria. This includes the 7 stars in which an accretion rate
could not be determined within the context of MA; it would require an
unrealistically high filling factor of shocked material to reproduce
their UV excess.  The measured $\Delta D_B$ of these 7 objects is
likely associated with accretion due to the overall properties of the stars, although the exact mechanism is not known (an alternative could be Boundary Layer accretion for these hot
objects \citet[see][]{Bertout1988, Blondel2006})\footnote{Since \macc{} was not
determined for these stars they do not contribute to the fits in
Figure \ref{fig:lacc_vs_lline}.}.

The two categories of emission and accretion can be divided into two outcomes for each case, those with and those without. This gives a total of four possible categories for a star to be in e.g. a star may have a particular line in emission, but no $\Delta D_B$ was detected in the star.

\begin{figure*}
\centering
\includegraphics[trim=0.5cm 0.25cm 0.0cm 1.0cm, width=\linewidth]{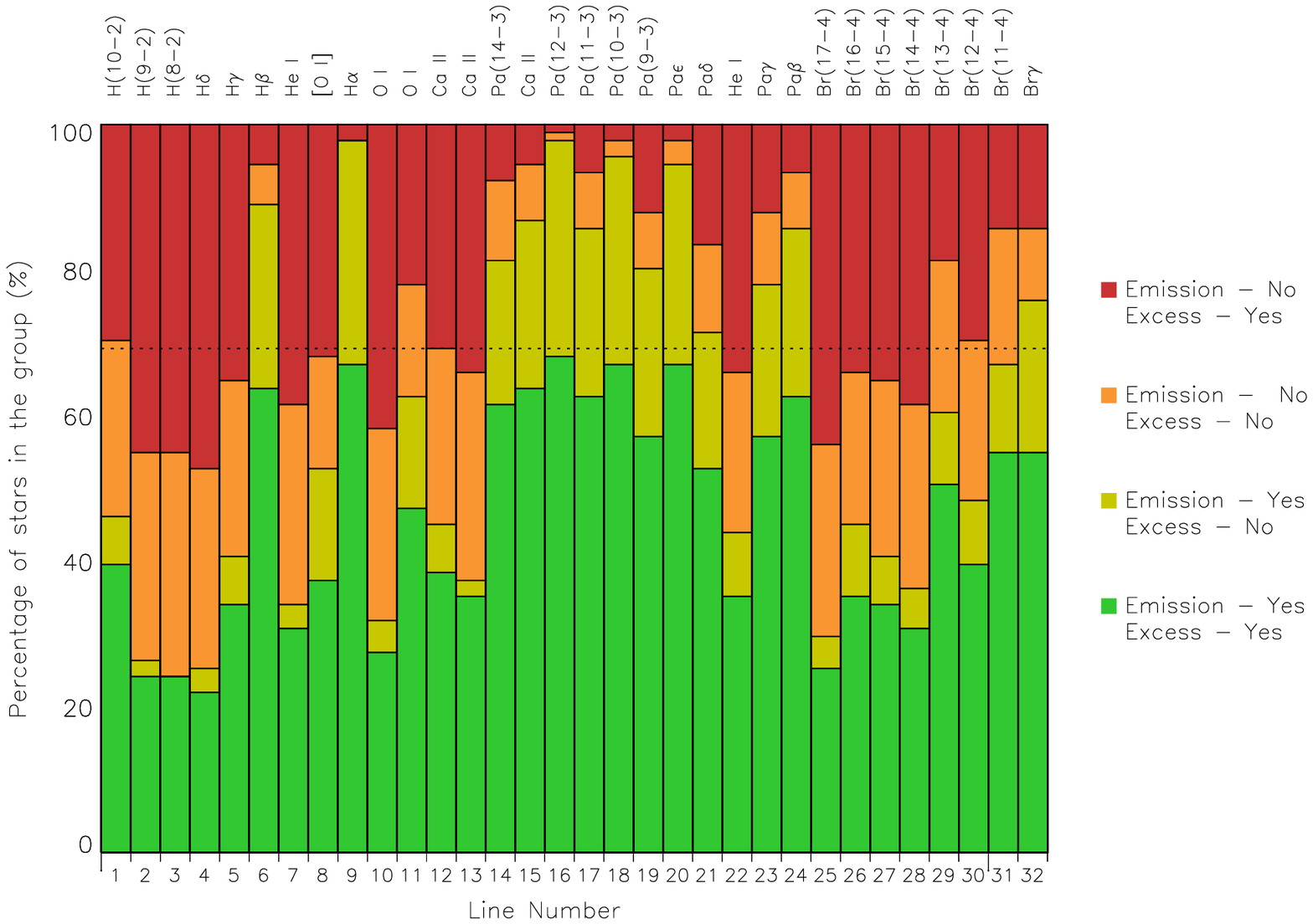}
\caption{The lines are shown by increasing wavelength along the xaxis, with their names given at the top of the plot. Each line is split into four different categories based on whether an emission line was/was not detected, and on whether $\Delta D_B$ could/could not be measured. All of the HAeBes examined are represented in this diagram. Details on the meanings of each category are provided in the text. The dash line denotes the percentage of the sample where an UV-excess is detected, 69\%, and therefore the closer the green region is to this line the better the tracer. A quick glance demonstrates that \halpha{}, \hbeta{}, and the majority of the Paschen series are all sensitive tracers, while both He I lines and weaker Brackett lines are less sensitive accretion diagnostic lines.}
\label{fig:percentage}
\end{figure*}

Table \ref{tab:yes_no}  gives the number of stars split into each category, for each line, and is presented visually in Figure \ref{fig:percentage}. What the four different categories represent are now detailed:
\begin{enumerate}

\item The first category contains objects for which both emission lines and a UV excess are
  detected, represented by the green segments in the
  diagram. This category measures if the line is a good one to one
  tracer of accretion i.e. excess is present and so is the emission
  line. Examples of good tracers are \halpha{},
  \brgamma{}, and \pabeta{} as their emission detections fall
  predominately in this category.

\item The second category indicates how difficult the line is to detect i.e. how insensitive it is. This category is shown in red and contains those stars for which
  a clear UV excess is detected and for which a mass accretion rate
  could be derived, but in whose spectra no emission line is
  detected. An example of this is the oxygen 7773\angstrom{} line, for
  which the majority of $\Delta D_B$ detections have no corresponding
  emission line detection.

\item The third category (shown in yellow in the figure) denotes the
  objects for which an emission line is detected, but no UV excess was
  detected. Assuming that the emission line is sensitive to accretion,
  and more readily detectable than an excess, the lines in this
  category allow the \lacc{} vs. \lline{} relationship to be used to
  infer an accretion luminosity.

\item The fourth and final category, shown in orange, contains those
  objects with no obvious UV excess and no emission in the line of
  interest. These objects generally have H$\alpha$ emission, but no
  sign of other emission lines. If they are actively accreting, the mass accretion
  rate is well below our detection limits (the lowest log(\macc{}) detected in \citetalias{Fairlamb2015} was only -7.78).

\end{enumerate}

The findings in the table and figure highlight that the already
established accretion tracing relationships are good ones; \halpha{},
\brgamma{}, and \pabeta{} are all predominately in the green.
There are also lines which
appear to be poor tracers of accretion. In particular the Brackett
series appears to get worse as an accretion tracer towards the higher
order transitions. Such a decline is likely because of the decreasing strength in
emission towards the shorter wavelengths; which is supported by the
green/red ratio changing between successive orders (with a slight
deviation around Br 12-4). The Balmer series demonstrates a similarity to this with the two lowest level transitions, \halpha{} and \hbeta{}, being sensitive to accretion while the remainder are less sensitive. The explanation for this is likely to be the same as for the Brackett series, however, another possible
explanation for this could be the UV-excess veiling the lines. Veiling of the lines occurs when their intrinsic absorption is filled in by excess emission, which
decreases the observed absorption in them and gives then an
\ewobs{} closer to 0. However, the brightest of HAeBes makes this scenario unlikely. 
The Paschen series is an exception to the other two hydrogen series; it maintains a constant level of sensitivity throughout the series. This is due to the location of the lines; the visible region. The SNR here is much higher than the UV and allows us to make a greater number of detections. While in the NIR, due to telluric line correction and IR-excess correction, there is greater margin of error for detection than the visible. As for the Ca II line it is likely that the Paschen blending results in more detections for the 8662 line than the other two, as it is blended with the stronger Paschen lines. 

As for lines which lean more towards being poor tracers of accretion, like the O I
7773\angstrom{} and the He I 5876\angstrom{} lines. A possible
explanation is that \ewint{} is measured as a sum over the entire line profile. This means that complex profiles, with multiple absorption and emission components, may be missing details of additional contributions to the lines. In extreme cases
stars may be found to have no emission detection despite the clear
presence of an inverse P-Cygni profile, where the absorption is strong enough to make the \ewint{} measurement uncertain. Therefore, when considering
accretion in HAeBes analysis of line profiles should be used in combination with line strengths and UV-excesses (where possible), to reach a consensus on detection and origin.
Such an example is presented by \citet{Cauley2014} where their analysis of the lines profiles of the He I 10830\micron{} line demonstrates strong evidence of MA occurring in HAes, and perhaps from HBes too. Despite the convincing line profiles the detection rate of the line being in overall emission is low, making it a poor tracer in that regard. However, if detected in emission it can still be used to infer \macc{}.

Despite information regarding line profiles being omitted in this work the \lacc{} - \lline{} relationships hold for the majority of stars and can be used as a diagnostic to infer accretion. Even the lines which have fewer detections correlate well with accretion and can be used if detected.

\subsection{Objects without UV excess}
\label{sec:results:no_uv_excess}

One particularly interesting group of stars are those that show lines in emission, but have no
detectable UV excess, $\Delta D_B$. This opens up the possibility that
the emission lines could be tracing accretion where the accretion rate is below the
detection limits of the Balmer excess method. This is now investigated here for the case of \halpha{}. 

The line luminosity of \halpha{} is converted into \lacc{},
using the relationship determined previously and tabulated in Table \ref{tab:lacc_lum}. For this test \lacc{} is
calculated fo the stars in which only an upper limit could be placed
on the Balmer excess (25 stars, see \citetalias{Fairlamb2015}). The derived \lacc{} is then converted into \macc{} using
the stellar parameters determined in \citetalias{Fairlamb2015}. This, in turn, is used to
predict the resulting $\Delta D_B$ values. The predicted excess can be
calculated via the $\Delta D_B$ vs. \macc{} curves seen in Figure 9 of \citetalias{Fairlamb2015}.

The predicted $\Delta D_B$ agrees with the upper limits measured in \citetalias{Fairlamb2015} for 21/25 of the objects. This agreement within previous limits for the majority of these stars suggests that the lines in this work are suitable as accretion tracers for inferring \macc{}. For the remaining 4 objects a $\Delta D_B$ is predicted that is higher than the upper limits previously measured; the values lie just 1$\sigma$ away from the previous limits. Considering that these fours results are only 1$\sigma$ away this is not significant to detract from the previous paragraph; the accretion diagnostic properties of the lines still holds in general.

Overall, we can conclude that the ``yes in emission, no in
excess'' criteria serves as a positive indicator that a line can be used as an accretion diagnostic line i.e. The line luminosity can be reliably used to infer the accretion rate.

\section{Conclusions}
\label{sec:disc_and_conc}

This study provides the largest spectroscopic investigation into accretion rates of HAeBes to date. In addition, the spectral wavelength range covered in these stars is much larger than any other HAeBe investigation, which allows an investigation from the UV up to the NIR at 2.5\micron{}. The combination of a large sample, 91 objects, and huge wavelength coverage allows for the most robust statistical investigation into the emission lines in HAeBes to be performed to date.

Line luminosities were obtained for 32 different lines. The
line luminosities are observed to be correlated with the accretion luminosities -- by
implication they are therefore correlated to the mass accretion rates. We focussed on expanding our
understanding of relationships between \lacc{} and \lline{} in HAeBes,
and how these relate to similar relationships observed in CTTs. The
large size of the sample and large spectral coverage resulted in a
tenfold increase in accretion line diagnostics for HAeBe stars
compared to what was known in the literature.

The following key points are found:

\begin{itemize}

\item Relationships are obtained between \lacc{} and \lline{} for 32
  different emission lines. In all cases a best fit is made to these
  lines, with the average correlation being $L_{\rm acc} \propto
  L_{\rm line}^{ 1.16 \pm 0.15 }$. The fact that both UV excess and
  emission lines were measured simultaneously make these the most robust set hitherto published.

 \item We find that all lines can be used as an accretion tracer. Additionally, it was shown by extrapolation, for the case of \halpha{}, that an
   emission line could be used to infer low mass accretion rates which cannot normally be measured
   by an UV excess. This is particularly applicable for the hotter objects
   where their large intrinsic UV output makes the detection of weak
   UV excess emission difficult.

\item The relationship $L_{\rm acc} \propto L_{\rm line}^{ 1.16 \pm 0.15}$ obtained for the HAeBes in this work agrees with the relationship observed in CTTs by \citet{Alcala2014}, where $L_{\rm acc} \propto L_{\rm line}^{ 1.08 \pm 0.08}$; the relationship is steeper for the HAeBes. On an individual line basis, some variations are seen in the relationship between the two. Notably the Ca and He lines,
  lines with complex line profiles that behave differently for both types
  of sources and warrants further investigation.

\item An assessment of the sensitivity of each line as an accretion
  tracer has been performed though a comparison of respective emission
  line and UV-excess detection rates. We confirm that the well-known
  accretion tracers for CTTs, such as \hbeta{}, \halpha{}, \pabeta{} and
  \brgamma{}, are also tracers of accretion in HAeBe
  stars. 

\item The sheer number of objects and emission lines analysed here
  provide robust relationships between \lacc{} and \lline{} for an exceptional wavelength coverage. In particular, known accretion tracing lines have been further verified and new observational windows have been opened up for measuring \macc{} in HAeBes e.g. \hbeta{} in the blue $B$-band region, \pabeta{} in the $J$-band region, and Br(13-4) in the $H$-band region. In the visible the Paschen series shows consistent sensitivity between lines. These new diagnostics will prove valuable in future observations.

\item Additionally, some lines can be considered poor tracers of accretion due to a low rate incidence of detection of emission lines e.g. \foi{} and He I 5876\angstrom{}. However, when these lines are
detected in emission they are nearly always associated with a UV-excess i.e. associated with accretion, and the relationship holds.
Since this aspect is present for all lines and relationships tested here, it is likely there is a deeper physical connection. For example, in the case of \foi{} it is suspected that photospheric UV photo-dissociates OH in upper layers of the disk, giving rise to this emission \citep{Acke2005}. 

\item The high number of correlations is a topic that has been discussed recently by \citet{Mendigutia2015}, where the authors found that the relationships observed between \lacc{} and \lline{} appear to be a consequence of them being directly related to the star's own stellar luminosity \citep[see also][]{Bohm1995}. Therefore, the line luminosities may not physically arise due to accretion onto the star, but they can still serve as a diagnostic for obtaining accretion rates.

\end{itemize}

In order to gain knowledge on the physical origin of the emission lines, and their association to accretion, the line profiles themselves must be considered. Work on line profiles so far has demonstrated notable differences between CTTs and HAeBes so far \citep{Hamann1992a, Hamann1992b, Cauley2014,Cauley2015}. Connecting these changes with both the relationships determined here, the underlying mathematical relationships \citep{Mendigutia2015}, and the large number of lines obtained with X-Shooter is the next challenge; this will will be presented in a future paper by the authors.

\section*{Acknowledgments}

JRF gratefully acknowledges a studentship from the Science and Technology Facilities Council of the UK. JDI gratefully acknowledges support from the DISCSIM project, grant agreement 341137, funded by the European Research Council under ERC-2013-ADG. This work has made use of NASA's Astrophysics Data System; it has also made use of the SIMBAD database, operated at CDS, Strasbourg, France. The authors would like to extend their thanks to the anonymous referee who helped strengthen the manuscript.

\bibliographystyle{mn2e}

\begin{thebibliography}{50}
\expandafter\ifx\csname natexlab\endcsname\relax\def\natexlab#1{#1}\fi

\bibitem[{{Acke}, {van den Ancker} \& {Dullemond}(2005){Acke}, {van den
  Ancker}, \& {Dullemond}}]{Acke2005}
{Acke} B., {van den Ancker} M.~E., {Dullemond} C.~P., 2005, \aap, 436, 209

\bibitem[{{Alcal{\'a}} {et~al}\mbox{.}(2014){Alcal{\'a}}, {Natta}, {Manara},
  {Spezzi}, {Stelzer}, {Frasca}, {Biazzo}, {Covino}, {Randich}, {Rigliaco},
  {Testi}, {Comer{\'o}n}, {Cupani}, \& {D'Elia}}]{Alcala2014}
{Alcal{\'a}} J.~M. {et~al.}, 2014, \aap, 561, A2

\bibitem[{{Bertout}(1989)}]{Bertout1989}
{Bertout} C., 1989, \araa, 27, 351

\bibitem[{{Bertout}, {Basri} \& {Bouvier}(1988){Bertout}, {Basri}, \&
  {Bouvier}}]{Bertout1988}
{Bertout} C., {Basri} G., {Bouvier} J., 1988, \apj, 330, 350

\bibitem[{{Blondel} \& {Djie}(2006)}]{Blondel2006}
{Blondel} P.~F.~C., {Djie} H.~R.~E.~T.~A., 2006, \aap, 456, 1045

\bibitem[{{Boehm} \& {Catala}(1995)}]{Bohm1995}
{Boehm} T., {Catala} C., 1995, \aap, 301, 155

\bibitem[{{Calvet} \& {Gullbring}(1998)}]{Calvet1998}
{Calvet} N., {Gullbring} E., 1998, \apj, 509, 802

\bibitem[{{Calvet} {et~al}\mbox{.}(2004){Calvet}, {Muzerolle}, {Brice{\~n}o},
  {Hern{\'a}ndez}, {Hartmann}, {Saucedo}, \& {Gordon}}]{Calvet2004}
{Calvet} N., {Muzerolle} J., {Brice{\~n}o} C., {Hern{\'a}ndez} J., {Hartmann}
  L., {Saucedo} J.~L., {Gordon} K.~D., 2004, \apj, 128, 1294

\bibitem[{{Castelli} \& {Kurucz}(2004)}]{Castelli2004}
{Castelli} F., {Kurucz} R.~L., 2004, {New Grids of ATLAS9 Model Atmospheres,
  ArXiv Astrophysics e-prints, arXiv:astro-ph/0405087}

\bibitem[{{Cauley} \& {Johns-Krull}(2014)}]{Cauley2014}
{Cauley} P.~W., {Johns-Krull} C.~M., 2014, \apj, 797, 112

\bibitem[{{Cauley} \& {Johns-Krull}(2015)}]{Cauley2015}
{Cauley} P.~W., {Johns-Krull} C.~M., 2015, \apj, 810, 5

\bibitem[{{Cutri} {et~al}\mbox{.}(2003){Cutri}, {Skrutskie}, {van Dyk},
  {Beichman}, {Carpenter}, {Chester}, {Cambresy}, {Evans}, {Fowler}, {Gizis},
  {Howard}, {Huchra}, {Jarrett}, {Kopan}, {Kirkpatrick}, {Light}, {Marsh},
  {McCallon}, {Schneider}, {Stiening}, {Sykes}, {Weinberg}, {Wheaton},
  {Wheelock}, \& {Zacarias}}]{Cutri2003}
{Cutri} R.~M. {et~al.}, 2003, {2MASS All Sky Catalog of point sources.}
  NASA/IPAC Infrared Science Archive

\bibitem[{{Dahm}(2008)}]{Dahm2008}
{Dahm} S.~E., 2008, \apj, 136, 521

\bibitem[{{Donehew} \& {Brittain}(2011)}]{Donehew2011}
{Donehew} B., {Brittain} S., 2011, \apj, 141, 46

\bibitem[{{Eiroa} {et~al}\mbox{.}(2001){Eiroa}, {Garz{\'o}n}, {Alberdi}, {de Winter}, {Ferlet}, {Grady}, {Collier Cameron}, {Davies}, {Deeg}, {Harris}, {Horne}, {Mer{\'{\i}}n}, {Miranda}, {Montesinos}, {Mora}, {Oudmaijer}, {Palacios},  {Penny}, {Quirrenbach}, {Rauer}, {Schneider}, {Solano}, {Tsapras}, {Wesselius}}]{Eiroa2001}{Eiroa} C. {et~al.}, 2001, \aap, 365, 110

\bibitem[{{Eiroa} {et~al}\mbox{.}(2002){Eiroa}, {Oudmaijer}, {Davies}, {de Winter}, {Garz{\'o}n}, {Palacios}, {Alberdi}, {Ferlet}, {Grady}, {Collier Cameron}, {Deeg}, {Harris}, {Horne}, {Mer{\'{\i}}n}, {Miranda}, {Montesinos}, {Mora}, {Penny}, {Quirrenbach}, {Rauer}, {Schneider}, {Solano}, {Tsapras}, {Wesselius}}]{Eiroa2002}{Eiroa}, C. {et~al.}, 2002, \aap, 384, 1038


\bibitem[{{Fairlamb} {et~al}\mbox{.}(2015){Fairlamb}, {Oudmaijer},
  {Mendigut{\'{\i}}a}, {Ilee}, \& {van den Ancker}}]{Fairlamb2015}
{Fairlamb} J.~R., {Oudmaijer} R.~D., {Mendigut{\'{\i}}a} I., {Ilee} J.~D., {van
  den Ancker} M.~E., 2015, \mnras, 453, 976

\bibitem[{{Finkenzeller} \& {Mundt}(1984)}]{Finkenzeller1984}
{Finkenzeller} U., {Mundt} R., 1984, \aaps, 55, 109

\bibitem[{{Garrison}(1978)}]{Garrison1978}
{Garrison}, Jr. L.~M., 1978, \apj, 224, 535

\bibitem[{{Ghosh} \& {Lamb}(1979)}]{Ghosh1979}
{Ghosh} P., {Lamb} F.~K., 1979, \apj, 232, 259

\bibitem[{{Gullbring} {et~al}\mbox{.}(2000){Gullbring}, {Calvet}, {Muzerolle},
  \& {Hartmann}}]{Gullbring2000}
{Gullbring} E., {Calvet} N., {Muzerolle} J., {Hartmann} L., 2000, \apj, 544,
  927

\bibitem[{{Gullbring} {et~al}\mbox{.}(1998){Gullbring}, {Hartmann}, {Briceno},
  \& {Calvet}}]{Gullbring1998}
{Gullbring} E., {Hartmann} L., {Briceno} C., {Calvet} N., 1998, \apj, 492, 323

\bibitem[{{Hamann} \& {Persson}(1992{\natexlab{a}})}]{Hamann1992a}
{Hamann} F., {Persson} S.~E., 1992{\natexlab{a}}, \apjs, 82, 247

\bibitem[{{Hamann} \& {Persson}(1992{\natexlab{b}})}]{Hamann1992b}
{Hamann} F., {Persson} S.~E., 1992{\natexlab{b}}, \apjs, 82, 285

\bibitem[{{Herbig}(1960)}]{Herbig1960}
{Herbig} G.~H., 1960, \apjs, 4, 337

\bibitem[{{Herczeg} \& {Hillenbrand}(2008)}]{Herczeg2008}
{Herczeg} G.~J., {Hillenbrand} L.~A., 2008, \apj, 681, 594

\bibitem[{{Ilee} {et~al}\mbox{.}(2014){Ilee}, {Fairlamb}, {Oudmaijer},
  {Mendigut{\'{\i}}a}, {van den Ancker}, {Kraus}, \& {Wheelwright}}]{Ilee2014}
{Ilee} J.~D., {Fairlamb} J., {Oudmaijer} R.~D., {Mendigut{\'{\i}}a} I., {van
  den Ancker} M.~E., {Kraus} S., {Wheelwright} H.~E., 2014, \mnras, 445, 3723

\bibitem[{{Ingleby} {et~al}\mbox{.}(2013){Ingleby}, {Calvet}, {Herczeg},
  {Blaty}, {Walter}, {Ardila}, {Alexander}, {Edwards}, {Espaillat}, {Gregory},
  {Hillenbrand}, \& {Brown}}]{Ingleby2013}
{Ingleby} L. {et~al.}, 2013, \apj, 767, 112

\bibitem[{{Kausch} {et~al}\mbox{.}(2015){Kausch}, {Noll}, {Smette},
  {Kimeswenger}, {Barden}, {Szyszka}, {Jones}, {Sana}, {Horst}, \&
  {Kerber}}]{Kausch2015}
{Kausch} W. {et~al.}, 2015, \aap, 576, A78

\bibitem[{{Koenigl}(1991)}]{Koenigl1991}
{Koenigl} A., 1991, \apjl, 370, L39

\bibitem[{{Kurucz}(1993)}]{Kurucz1993}
{Kurucz} R.~L., 1993, {SYNTHE spectrum synthesis programs and line data}.
  Cambridge, MA: Smithsonian Astrophysical Observatory

\bibitem[{{Meeus} {et~al}\mbox{.}(2001){Meeus}, {Waters}, {Bouwman}, {van den
  Ancker}, {Waelkens}, \& {Malfait}}]{Meeus2001}
{Meeus} G., {Waters} L.~B.~F.~M., {Bouwman} J., {van den Ancker} M.~E.,
  {Waelkens} C., {Malfait} K., 2001, \aap, 365, 476

\bibitem[{{Mendigut{\'{i}}a} {et~al}\mbox{.}(2011){Mendigut{\'{i}}a}, {Calvet},
  {Montesinos}, {Mora}, {Muzerolle}, {Eiroa}, {Oudmaijer}, \&
  {Mer{\'{\i}}n}}]{Mendigutia2011b}
{Mendigut{\'{i}}a} I., {Calvet} N., {Montesinos} B., {Mora} A., {Muzerolle} J.,
  {Eiroa} C., {Oudmaijer} R.~D., {Mer{\'{\i}}n} B., 2011, \aap, 535, A99

\bibitem[{{Mendigut{\'{i}}a} {et~al}\mbox{.}(2013){Mendigut{\'{i}}a},
  {Brittain}, {Eiroa}, {Meeus}, {Montesinos}, {Mora}, {Muzerolle}, {Oudmaijer},
  \& {Rigliaco}}]{Mendigutia2013}
{Mendigut{\'{i}}a} I. {et~al.}, 2013, \apj, 776, 44

\bibitem[{{Mendigut{\'{\i}}a} {et~al}\mbox{.}(2015){Mendigut{\'{\i}}a},
  {Oudmaijer}, {Rigliaco}, {Fairlamb}, {Calvet}, {Muzerolle}, {Cunningham}, \&
  {Lumsden}}]{Mendigutia2015}
{Mendigut{\'{\i}}a} I., {Oudmaijer} R.~D., {Rigliaco} E., {Fairlamb} J.~R.,
  {Calvet} N., {Muzerolle} J., {Cunningham} N., {Lumsden} S.~L., 2015, \mnras,
  452, 2837

\bibitem[{{Mottram} {et~al}\mbox{.}(2011){Mottram}, {Hoare}, {Urquhart},
  {Lumsden}, {Oudmaijer}, {Robitaille}, {Moore}, {Davies}, \&
  {Stead}}]{Mottram2011}
{Mottram} J.~C. {et~al.}, 2011, \aap, 525, A149

\bibitem[{{Munari} {et~al}\mbox{.}(2005){Munari}, {Sordo}, {Castelli}, \&
  {Zwitter}}]{Munari2005}
{Munari} U., {Sordo} R., {Castelli} F., {Zwitter} T., 2005, \aap, 442, 1127

\bibitem[{{Muzerolle}, {Hartmann} \& {Calvet}(1998){Muzerolle}, {Hartmann}, \&
  {Calvet}}]{Muzerolle1998b}
{Muzerolle} J., {Hartmann} L., {Calvet} N., 1998, \apj, 116, 455

\bibitem[{{Muzerolle} {et~al}\mbox{.}(2004){Muzerolle}, {D'Alessio}, {Calvet},
  \& {Hartmann}}]{Muzerolle2004}
{Muzerolle} J., {D'Alessio} P., {Calvet} N., {Hartmann} L., 2004, \apj, 617,
  406

\bibitem[{{Rigliaco} {et~al}\mbox{.}(2012){Rigliaco}, {Natta}, {Testi},
  {Randich}, {Alcal{\`a}}, {Covino}, \& {Stelzer}}]{Rigliaco2012}
{Rigliaco} E., {Natta} A., {Testi} L., {Randich} S., {Alcal{\`a}} J.~M.,
  {Covino} E., {Stelzer} B., 2012, \aap, 548, A56

\bibitem[{{Shu} {et~al}\mbox{.}(1994){Shu}, {Najita}, {Ostriker}, {Wilkin},
  {Ruden}, \& {Lizano}}]{Shu1994}
{Shu} F., {Najita} J., {Ostriker} E., {Wilkin} F., {Ruden} S., {Lizano} S.,
  1994, \apj, 429, 781

\bibitem[{{Skrutskie} {et~al}\mbox{.}(2006){Skrutskie}, {Cutri}, {Stiening},
  {Weinberg}, {Schneider}, {Carpenter}, {Beichman}, {Capps}, {Chester},
  {Elias}, {Huchra}, {Liebert}, {Lonsdale}, {Monet}, {Price}, {Seitzer},
  {Jarrett}, {Kirkpatrick}, {Gizis}, {Howard}, {Evans}, {Fowler}, {Fullmer},
  {Hurt}, {Light}, {Kopan}, {Marsh}, {McCallon}, {Tam}, {Van Dyk}, \&
  {Wheelock}}]{Skrutskie2006}
{Skrutskie} M.~F. {et~al.}, 2006, \apj, 131, 1163

\bibitem[{{Smette} {et~al}\mbox{.}(2015){Smette}, {Sana}, {Noll}, {Horst},
  {Kausch}, {Kimeswenger}, {Barden}, {Szyszka}, {Jones}, {Gallenne}, {Vinther},
  {Ballester}, \& {Taylor}}]{Smette2015}
{Smette} A. {et~al.}, 2015, \aap, 576, A77

\bibitem[{{Th{\'{e}}}, {de Winter} \& {Perez}(1994){Th{\'{e}}}, {de Winter}, \&
  {Perez}}]{The1994}
{Th{\'{e}}} P.~S., {de Winter} D., {Perez} M.~R., 1994, \aaps, 104, 315

\bibitem[{{Uchida} \& {Shibata}(1985)}]{Uchida1985}
{Uchida} Y., {Shibata} K., 1985, \pasj, 37, 515

\bibitem[{{van den Ancker} {et~al}\mbox{.}(2000){van den Ancker}, {Bouwman},
  {Wesselius}, {Waters}, {Dougherty}, \& {van Dishoeck}}]{vandenAncker2000}
{van den Ancker} M.~E., {Bouwman} J., {Wesselius} P.~R., {Waters} L.~B.~F.~M.,
  {Dougherty} S.~M., {van Dishoeck} E.~F., 2000, \aap, 357, 325

\bibitem[{{Vernet} {et~al}\mbox{.}(2011){Vernet}, {Dekker}, {D'Odorico},
  {Kaper}, {Kjaergaard}, {Hammer}, {Randich}, {Zerbi}, {Groot}, {Hjorth},
  {Guinouard}, {Navarro}, {Adolfse}, {Albers}, {Amans}, {Andersen}, {Andersen},
  {Binetruy}, {Bristow}, {Castillo}, {Chemla}, {Christensen}, {Conconi},
  {Conzelmann}, {Dam}, {de Caprio}, {de Ugarte Postigo}, {Delabre}, {di
  Marcantonio}, {Downing}, {Elswijk}, {Finger}, {Fischer}, {Flores}, {Fran{\c
  c}ois}, {Goldoni}, {Guglielmi}, {Haigron}, {Hanenburg}, {Hendriks},
  {Horrobin}, {Horville}, {Jessen}, {Kerber}, {Kern}, {Kiekebusch}, {Kleszcz},
  {Klougart}, {Kragt}, {Larsen}, {Lizon}, {Lucuix}, {Mainieri}, {Manuputy},
  {Martayan}, {Mason}, {Mazzoleni}, {Michaelsen}, {Modigliani}, {Moehler},
  {M{\o}ller}, {Norup S{\o}rensen}, {N{\o}rregaard}, {P{\'e}roux}, {Patat},
  {Pena}, {Pragt}, {Reinero}, {Rigal}, {Riva}, {Roelfsema}, {Royer}, {Sacco},
  {Santin}, {Schoenmaker}, {Spano}, {Sweers}, {Ter Horst}, {Tintori}, {Tromp},
  {van Dael}, {van der Vliet}, {Venema}, {Vidali}, {Vinther}, {Vola},
  {Winters}, {Wistisen}, {Wulterkens}, \& {Zacchei}}]{Vernet2011}
{Vernet} J. {et~al.}, 2011, \aap, 536, A105

\bibitem[{{Vieira} {et~al}\mbox{.}(2003){Vieira}, {Corradi}, {Alencar},
  {Mendes}, {Torres}, {Quast}, {Guimar{\~a}es}, \& {da Silva}}]{Vieira2003}
{Vieira} S.~L.~A., {Corradi} W.~J.~B., {Alencar} S.~H.~P., {Mendes} L.~T.~S.,
  {Torres} C.~A.~O., {Quast} G.~R., {Guimar{\~a}es} M.~M., {da Silva} L., 2003,
  \apj, 126, 2971

\bibitem[{{Vink} {et~al}\mbox{.}(2002){Vink}, {Drew}, {Harries}, \&
  {Oudmaijer}}]{Vink2002}
{Vink} J.~S., {Drew} J.~E., {Harries} T.~J., {Oudmaijer} R.~D., 2002, \mnras,
  337, 356

\bibitem[{{Vink} {et~al}\mbox{.}(2005){Vink}, {Drew}, {Harries}, {Oudmaijer},
  \& {Unruh}}]{Vink2005}
{Vink} J.~S., {Drew} J.~E., {Harries} T.~J., {Oudmaijer} R.~D., {Unruh} Y.,
  2005, \mnras, 359, 1049

\end{thebibliography}

\newcommand{\noop}[1]{}

\appendix

\section{Additional Luminosity Relationships}

The sheer number of lines analysed here warrants the use of an appendix for easy reading. In here, Figure \ref{fig:lacc_vs_lline_append} displays the accretion luminosity versus line luminosity relationships for all 32 lines analysed (an extension of Figure \ref{fig:lacc_vs_lline}). The best-fits to the data are present in the figure. Where possible the data on CTTs from \citet{Alcala2014} is also included for comparison.


\begin{figure*}
\centering
\includegraphics[trim=0.4cm 0.4cm 0.4cm 0.4cm, width=0.95\textwidth]{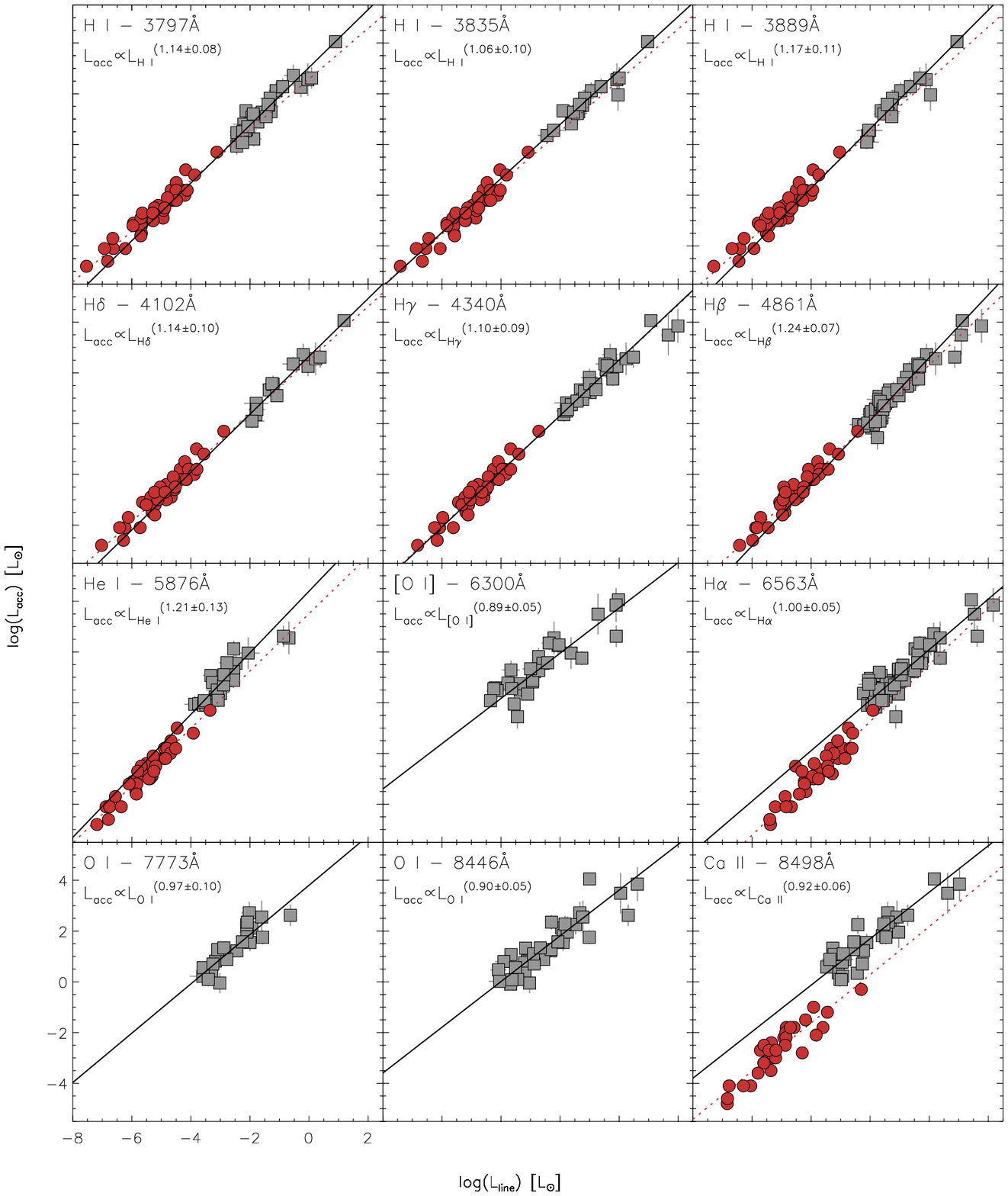}
\caption{ This figure shows the accretion luminosity, as calculated directly from the UV-excess, against the line luminosities.  Each panel shows only the HAeBes where both a clear UV-excess and clear emission line detection are made, shown as grey squares;a best-fit to them is shown as a solid-black line (the equation for the line is given below the line descriptor in the top-left of each panel, and is also given in \ref{tab:lacc_lum}). Where applicable, the data on CTTs from \citet{Alcala2014} is plotted for comparison, as red circles, along with their best-fit as a dashed-red line.}
\label{fig:lacc_vs_lline_append}
\end{figure*}

\begin{figure*}\ContinuedFloat
\centering
\includegraphics[trim=0.4cm 0.4cm 0.4cm 0.4cm, width=0.95\textwidth]{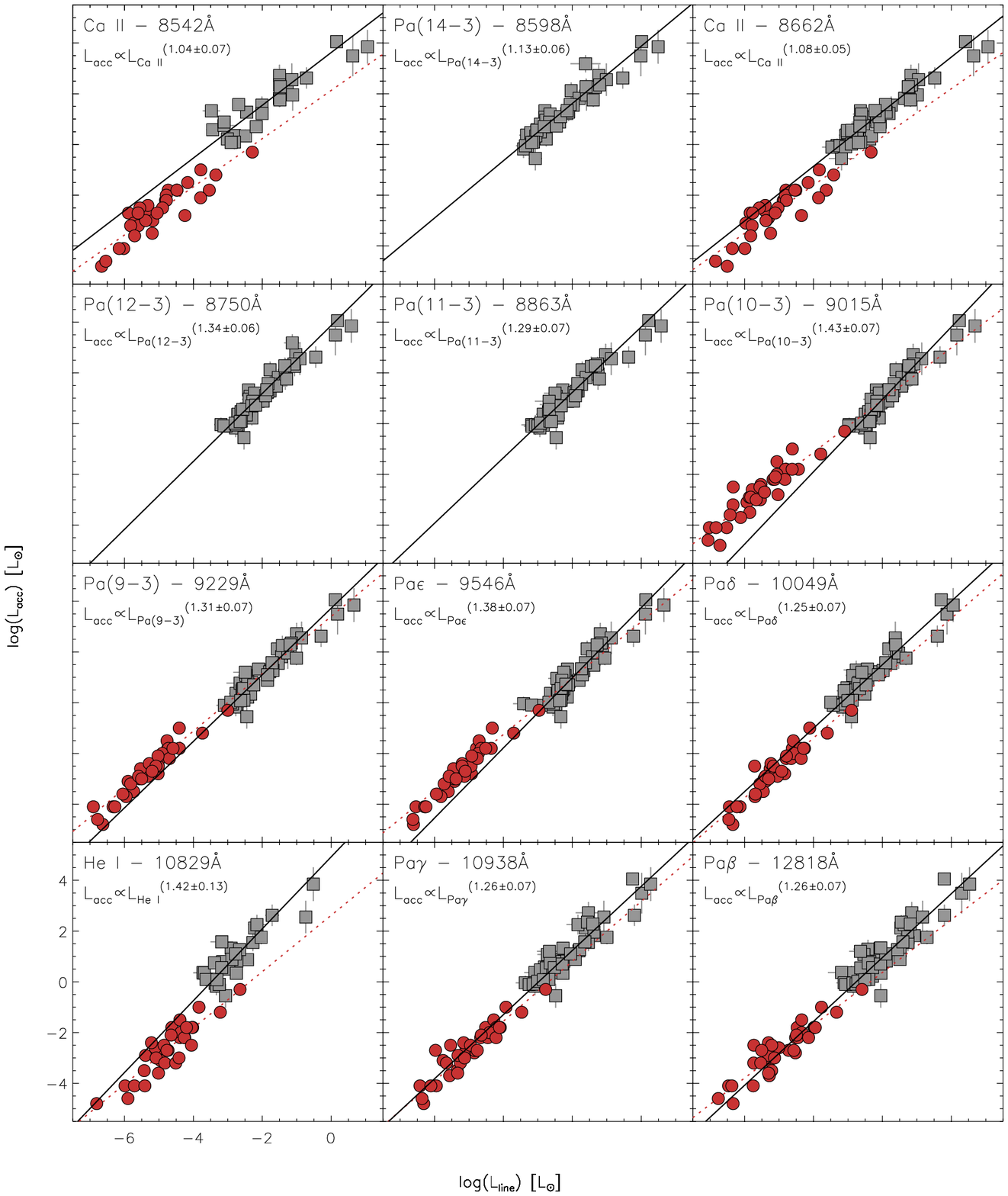}
\caption{continued.}
\end{figure*}

\begin{figure*}\ContinuedFloat
\centering
\includegraphics[trim=0.4cm 4.0cm 0.4cm 0.4cm, width=0.95\textwidth]{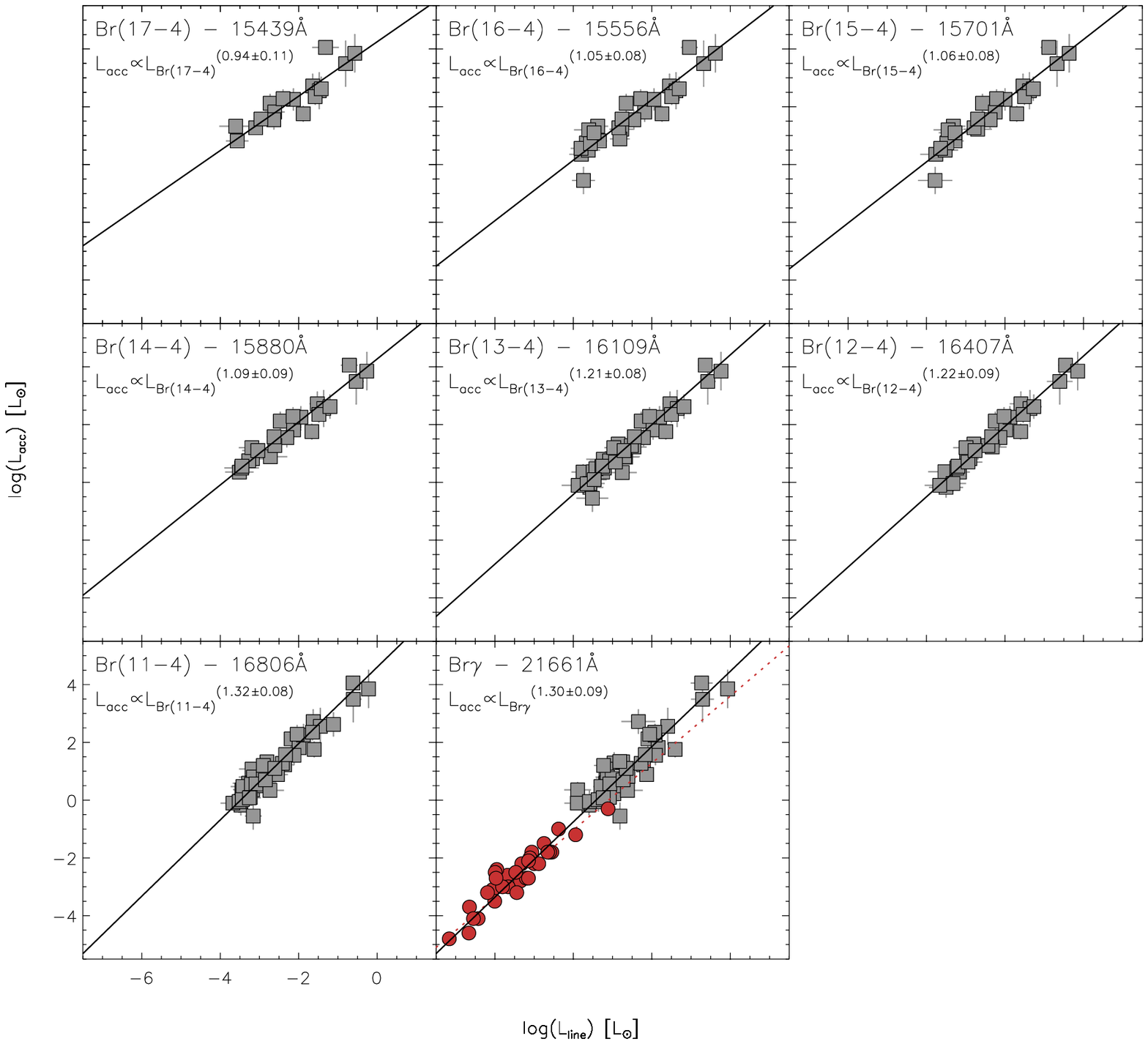}
\caption{continued.}
\end{figure*}


\label{lastpage}

\end{document}